\documentclass[11pt]{article}

\usepackage{a4wide}
\usepackage{amsmath}
\usepackage{amsthm}
\usepackage{amsfonts}
\usepackage{amssymb}
\usepackage{graphicx}

\makeatletter
\@addtoreset{equation}{section}

\makeatother

\def\e{\mathrm{e}}

\def\i{\sqrt{-1}}
\def\Define{:=}
\def\definE{=:}
\newcommand{\vecvar}[1]{\mbox{\boldmath $#1$}}
\def\nn{\nonumber}

\title
{\Large\textbf{
Bethe-ansatz studies of energy level crossings\\[2mm]
in the one-dimensional Hubbard model
}}

\author{
Akinori Nishino\footnote{E-mail address: %
nishino@gokutan.c.u-tokyo.ac.jp}
and
Tetsuo Deguchi$^{1}$\footnote{E-mail address: %
deguchi@phys.ocha.ac.jp}\\
\,\\
Institute of Physics, University of Tokyo, \\[2mm]
3--8--1 Komaba, Meguro-Ku, Tokyo 153--8902, Japan \\
\,\\
$^{1}$Department of Physics, Ochanomizu University,\\[2mm]
2--1--1 Otsuka, Bunkyo-ku, Tokyo, 112--8610, Japan
}

\date{}

\begin{document}

\setlength{\baselineskip}{18pt}

\maketitle

\noindent
\begin{abstract}
\setlength{\baselineskip}{15pt}
Motivated by Heilmann and Lieb's work~\cite{Heilmann-Lieb_71}, 
we discuss energy level crossings for the one-dimensional Hubbard model 
through the Bethe ansatz, constructing explicitly 
the degenerate eigenstates at the crossing points. 
After showing the existence of solutions for the Lieb--Wu equations 
with one-down spin, we solve them numerically and construct 
Bethe ansatz eigenstates.  
We thus verify all the level crossings in the spectral flows  
observed by the numerical diagonalization method with one down-spin. 
For each of  the solutions we obtain its energy spectral 
flow along the interaction parameter $U$. Then, 
we observe that some of the energy level crossings  can not 
be explained in terms of $U$-independent symmetries. 
Dynamical symmetries of the Hubbard model are fundamental for identifying  
each of the spectral lines at the level crossing points.   
We show that the Bethe ansatz eigenstates which 
degenerate at the points have distinct sets of eigenvalues 
of the higher conserved operators.  
We also show a twofold  permanent degeneracy 
in terms of the Bethe ansatz wavefunction. 
\end{abstract}

\section{Introduction}

Degeneracies in the energy spectra of quantum systems 
have close relationships with their symmetries. 
Actually, von Neumann and Wigner showed that 
degeneracies are more likely to occur for the systems 
with one or more symmetries than those without 
symmetries~\cite{Neumann-Wigner_29PZ,Reichl}. 
To be more precise, 
if one assumes that a quantum system is given by 
 a real Hamiltonian matrix  
whose elements are expressed by independent parameters, 
in the case of no symmetry, 
two parameters happen to take some prescribed values 
in order to bring two of the eigenvalues 
into coincidence. 
Their theory reminds us of the ``non-crossing rule''
in quantum chemistry, which states that energy levels of 
orbitals of the same symmetry can never cross 
each other along a reaction parameter. However, 
von Neumann--Wigner's theorem does not give a proof 
for the non-crossing rule. It is possible that  
degeneracies appear in the systems without symmetries.  
Such degeneracies are referred to as accidental degeneracies. 
In fact some examples of accidental degeneracies 
are numerically observed in molecules~\cite{vazPires_JPC78}
or triangular quantum billiards~\cite{Berry-Wilkinson_84PRSL}.

\par 
The one-dimensional Hubbard model is one of the most significant 
 models in condensed matter physics. 
The model also attracts a great interest of mathematical physicists
due to its Bethe-ansatz solvability~\cite{Lieb-Wu_68PRL,Korepin-Essler}.
Heilmann and Lieb numerically investigated energy spectral flows
along the interaction parameter $U$ 
for the system on a periodic 6-site chain and
found many level crossings which can not be accounted for 
by the known symmetries such as translation, 
$SO(4)$ and particle-hole 
symmetries~\cite{Heilmann-Lieb_71,Bondeson-Soos_79JCP}.
They concluded that, if one takes into account only 
$U$-independent symmetries,
the level crossings are accidental degeneracies,
that is, a counter example of the non-crossing rule.
Recently, Yuzbashyan, Altshuler and Shastry have suggested that 
the origin of Heilmann--Lieb's level crossings 
should be 
dynamical symmetries in the Hubbard model~\cite{Shastry_02JPA}.
Here the dynamical symmetries are given by 
parameter-dependent operators, which are often called higher 
conserved operators in association with conserved quantities 
in classical integrable systems. 
The dynamical symmetries for the Hubbard model are 
constructed in~\cite{Shastry_86PRL,Shastry_88JSP,Grosse_89LMP,%
Wadati-Olmedilla_87JPSJ,Olmedilla-Wadati_88PRL}.
Yuzbashyan, Altshuler and Shastry numerically showed that crossings 
in the spectral flows of the first three conserved operators 
never occur at the same value of $U$. 
The dynamical symmetries depend on the parameter $U$, 
and they are not considered as symmetries 
 in the von Neumann--Wigner's theorem.  
 Heilmann--Lieb's level crossings are still considered to 
be accidental degeneracies.

\par
In the framework of the Bethe ansatz,  
we discuss in the paper 
energy level crossings for the one-dimensional Hubbard model. 
The Bethe ansatz method provides information 
on the eigenstates that can not be 
easily obtained in the direct diagonalization of the Hamiltonian matrix.
We set up the following problems:
\begin{enumerate}
\item[i)] When two energy eigenvalues approach 
in numerical data as one parameter is varied,  
one may draw two alternative spectral flows, 
a level crossing or level repulsion~\cite{Heilmann-Lieb_71}.
To ensure that genuine energy level crossings have happened, 
we must investigate the change of each eigenstate 
along the parameter $U$. 
\item[ii)] Do the eigenstates have distinct  dynamical symmetries  
at the level crossing points? 
In order to solve the problem 
we assign  each of the eigenstates 
 the eigenvalues of the higher conserved operators.
\end{enumerate}
It is indeed not easy to investigate these problems.  
For the triangular quantum billiards with two parameters, 
Berry and Wilkinson investigated the behaviour of eigenstates 
along a circuit of the crossing point in the parameter space
so that they could verify the existence of genuine energy 
level crossings~\cite{Berry-Wilkinson_84PRSL}.
For general quantum systems,  
it is hard in practice 
to construct quantum many-body eigenstates. 
The task is also not easy even for the systems 
that can be treated by the Bethe ansatz. 
In fact, it is nontrivial to obtain  numerical solutions to 
the Bethe ansatz equations for a finite size system.  
For the sector of one down-spin, however, 
it is practically 
possible to solve the Bethe ansatz equations numerically. 

\par 
In the present paper, 
we prove the existence of solutions to the Lieb--Wu equations 
and then  numerically solve them. Here we generalize the method 
of Ref.~\cite{Deguchi_00PR}. 
By using the numerical solutions, 
we analyse the behaviour of each of the degenerate eigenstates 
along the parameter $U$. 
We thus find several genuine level crossings of energy spectral flows 
with the same $U$-independent symmetries.
The merit of this method is  that the numerical solutions 
for the Lieb--Wu equations provide not only the eigenvalues of the higher 
conserved operators but also the one-to-one correspondence 
between their spectral lines and eigenstates. 
As a consequence, we observe that all the common 
eigenspaces of the first three higher conserved operators 
are one-dimensional in the subspaces 
with the same $U$-independent symmetries. 
Furthermore, when there 
are $U$-independent degeneracies (permanent degeneracies) 
in the spectral flow, the one-to-one correspondence 
plays an essential role in assigning to each of the degenerate 
eigenstates  its eigenvalues of 
the dynamical symmetries at the energy level crossing points. 
By using the explicit form of 
the Bethe ansatz wavefunctions~\cite{Woynarovich_82JPC}, 
we can derive the permanent degeneracies.  
We remark that 
the existence of level crossings also gives 
a necessary condition for algebraic independence of 
the three higher conserved operators in the subspaces. 

\par 
This paper is organized as follows: 
first, in Section~\ref{sec:symmetry-Bethe},
we summarize the known results for the Hubbard model and 
prepare for the following sections.
We list the known symmetries of the Hubbard model 
in~\ref{sec:symmetries} and review the Bethe ansatz method in~\ref{sec:BAM}.
In~\ref{sec:LW-eq_one-down} we prove the existence of solutions
for the Lieb--Wu equations with one down-spin 
following the approach developed in \cite{Deguchi_00PR}.
Next, in Section~\ref{sec:crossings},
we analyse degeneracies in the energy spectrum
of the Hubbard model by using both direct diagonalization
of the Hamiltonian matrix and the Bethe ansatz method.
In~\ref{sec:twofold} we describe twofold permanent degeneracies
arising from a reflection symmetry of the lattice.
In~\ref{sec:631} and \ref{sec:641} we investigate
energy level crossings for the systems with two or three 
up-spins and one down-spin on a periodic 6-site chain.
We find some genuine energy level crossings and
see that the degenerate eigenstates can be classified 
by the eigenvalues of the higher conserved operators.
The final section is devoted to concluding remarks.

\section{Symmetries and Bethe ansatz method}
\label{sec:symmetry-Bethe}

We introduce the Hubbard model on a 
one-dimensional periodic $L$-site chain.
Let $c_{is}^{\dagger}$ and $c_{is}, 
(i\in\mathbb{Z}/L\mathbb{Z}, 
s\in\{\uparrow,\downarrow\})$ be
the creation and annihilation operators of electrons satisfying
$\{c_{is},c_{jt}\}=\{c_{is}^{\dagger},c_{jt}^{\dagger}\}=0$ and
$\{c_{is},c_{jt}^{\dagger}\}=\delta_{ij}\delta_{st}$,
and define the number operators by 
$n_{is}\Define c_{is}^{\dagger}c_{is}$.
We consider the Fock space of electrons 
with the vacuum state $|0\rangle$.
The one-dimensional Hubbard model is described by the following 
Hamiltonian acting on the Fock space:
\begin{align}
\label{eq:Hubbard_Ham}
 &H=-\sum_{i=1}^{L}\sum_{s=\uparrow,\downarrow}
  (c_{is}^{\dagger}c_{i+1,s}+c_{i+1,s}^{\dagger}c_{is})
  +U\sum_{i=1}^{L}\Big(n_{i\uparrow}-\frac{1}{2}\Big)
                  \Big(n_{i\downarrow}-\frac{1}{2}\Big),
\end{align}
where $U(\in\mathbb{R}_{>0})$ is the interaction parameter.
We consider the system for $L$ even throughout the article.

\subsection{$U$-independent and dynamical symmetries}
\label{sec:symmetries}

In general, the symmetries of a quantum system are expressed by 
operators which commute with its Hamiltonian.
They are classified into two families in the case of the Hubbard model;
one is independent of $U$ and another depends on $U$.
We list some of these symmetries.
First we consider the $U$-independent 
symmetries~\cite{Heilmann-Lieb_71}.
Define 
\begin{align}
 \sigma\Define
  \prod_{i=1}^{\frac{L}{2}-1}\prod_{s=\uparrow,\downarrow}
  P^{(s)}_{i,L-i}, \quad
  \tau\Define\prod_{s=\uparrow,\downarrow}
  P^{(s)}_{12}P^{(s)}_{23}\cdots P^{(s)}_{L-1,L}, 
\end{align}
where $
  P^{(s)}_{ij}\Define 
  1-(c_{is}^{\dagger}-c_{js}^{\dagger})(c_{is}-c_{js}),
  (i\neq j)$
are permutation operators.
The operators $\sigma$ and $\tau$ correspond to reflection and 
translation symmetries of the lattice, respectively.
They satisfy $\sigma^{2}=\tau^{L}=1$.
It is clear that
they commute with the Hamiltonian~\eqref{eq:Hubbard_Ham}.
Another $U$-independent symmetry is the $SO(4)$ 
symmetry~\cite{Yang_89PRL,Yang-Zhang_90MPL}. Define
\begin{align}
\label{eq:so(4)}
 &S_{z}\Define
  \frac{1}{2}\sum_{i=1}^{L}(n_{i\uparrow}-n_{i\downarrow}),\quad
  S_{+}\Define
  \sum_{i=1}^{L}c_{i\uparrow}^{\dagger}c_{i\downarrow},\quad
  S_{-}\Define(S_{+})^{\dagger}, \nn\\
 &\eta_{z}\Define
  \frac{1}{2}\sum_{i=1}^{L}(1-n_{i\uparrow}-n_{i\downarrow}),\quad
  \eta_{+}\Define
  \sum_{i=1}^{L}(-)^{i}
  c_{i\downarrow}c_{i\uparrow},\quad
  \eta_{-}\Define(\eta_{+})^{\dagger}.
\end{align}
Both sets of operators $\{S_{z},S_{\pm}\}$ and $\{\eta_{z},\eta_{\pm}\}$
give representations of the algebra $\mathfrak{su}(2)$ in the Fock space.
They all commute with the Hamiltonian~\eqref{eq:Hubbard_Ham}, 
which leads to the symmetry of type
$\mathfrak{su}(2)\oplus\mathfrak{su}(2)\simeq\mathfrak{so}(4)$.
Furthermore it is known that 
this $\mathfrak{so}(4)$ symmetry
lifts to the $SO(4)$ group symmetry.
For the later discussion, we define Casimir operators,
\begin{align}
 \vecvar{S}^{2}
  \Define\frac{1}{2}\Big(S_{+}S_{-}+S_{-}S_{+}\Big)+S_{z}^{2},\quad
 \vecvar{\eta}^{2}
  \Define\frac{1}{2}\Big(\eta_{+}\eta_{-}+\eta_{-}\eta_{+}\Big)
  +\eta_{z}^{2},\nn
\end{align}
which commute with all the operators in~\eqref{eq:so(4)}.

Next we introduce the dynamical symmetries given by 
the $U$-dependent operators.
These $U$-dependent operators themselves are also commutative 
and are called conserved operators in association with conserved
quantities in classical integrable systems.
In~\cite{Shastry_88JSP,Grosse_89LMP},
the first three conserved operators are explicitly given by
\begin{align}
 I_{1}=&H, \nn\\
 I_{2}
 =&-\i\sum_{i=1}^{L}\sum_{s=\uparrow,\downarrow}
  (c_{is}^{\dagger}c_{i+2,s}\!-\!c_{i+2,s}^{\dagger}c_{is}) \nn\\
  &\!+\!\i U\sum_{i=1}^{L}\sum_{s=\uparrow,\downarrow}
  (c_{is}^{\dagger}c_{i+1,s}\!-\!c_{i+1,s}^{\dagger}c_{is})
  (n_{i,-s}\!+\!n_{i+1,-s}\!-\!1), \nn\\
 I_{3}=
 &-\sum_{i=1}^{L}\sum_{s=\uparrow,\downarrow}
  (c_{is}^{\dagger}c_{i+3,s}+c_{i+3,s}^{\dagger}c_{is}) \nn\\
 &+U\sum_{i=1}^{L}\!\sum_{s=\uparrow,\downarrow}
  \Big((c_{is}^{\dagger}c_{i+2,s}+c_{i+2,s}^{\dagger}c_{is})
  \Big(n_{i,-s}+n_{i+1,-s}+n_{i+2,-s}-\frac{3}{2}\Big) \nn\\
 &\hspace{30mm} 
  +(c_{i+1,s}^{\dagger}c_{i+2,s}\!-\!c_{i+2,s}^{\dagger}c_{i+1,s})
  (c_{i,-s}^{\dagger}c_{i+1,-s}\!-\!c_{i+1,-s}^{\dagger}c_{i,-s}) \nn\\
 &\hspace{30mm} 
  +\frac{1}{2}
  (c_{i,s}^{\dagger}c_{i+1,s}\!-\!c_{i+1,s}^{\dagger}c_{i,s})
  (c_{i,-s}^{\dagger}c_{i+1,-s}\!-\!c_{i+1,-s}^{\dagger}c_{i,-s}) \nn\\
 &\hspace{30mm} 
  -\Big(n_{is}\!-\!\frac{1}{2}\Big)
  \Big(n_{i+1,-s}\!-\!\frac{1}{2}\Big)
  -\frac{1}{2}
  \Big(n_{is}\!-\!\frac{1}{2}\Big)
  \Big(n_{i,-s}\!-\!\frac{1}{2}\Big)\Big) \nn\\
 &-U^{2}\sum_{i=1}^{L}\!\sum_{s=\uparrow,\downarrow}
  (c_{is}^{\dagger}c_{i+1,s}+c_{i+1,s}^{\dagger}c_{is})
  \Big(n_{i,-s}n_{i+1,-s}\!-\!\frac{1}{2}(n_{i,-s}\!+\!n_{i+1,-s})
       \!+\!\frac{1}{2}\Big).
 \nn
\end{align}
In order to obtain higher conserved operators systematically,
the transfer matrix approach similar to 
that of the XXZ Heisenberg spin chain is 
developed~\cite{Shastry_86PRL,Shastry_88JSP,%
Wadati-Olmedilla_87JPSJ,Olmedilla-Wadati_88PRL}.
The $SO(4)$ symmetry of such higher conserved operators 
is also verified in the framework of the transfer matrix 
approach~\cite{Shiroishi-Ujino_98JPA}.

The operators $\{\tau,\vecvar{S}^{2},S_{z},
\vecvar{\eta}^{2},\eta_{z},I_{n}\}$ give a commutative set of 
operators including the Hamiltonian.  
Hence they can be diagonalized simultaneously.

\subsection{Bethe ansatz method}
\label{sec:BAM}

The Bethe ansatz method was applied to the Hubbard model
in~\cite{Lieb-Wu_68PRL}.
Here we give only the result.
Let $N$ be the number of electrons and $M$ that of down-spins.
We may assume $2M\leq N\leq L$ due to particle-hole
and spin reversal symmetries in the system.
Let $k=\{k_{i}|i=1,2,\ldots,N\}, 
(\mathrm{Re}(k_{i})\in\mathbb{R}/2\pi\mathbb{R})$
denote a set of wavenumbers of electrons and
$\lambda=\{\lambda_{\alpha}|\alpha=1,2,\ldots,M\}$
that of rapidities of down-spins.
Given a set of spin configuration
$s=\{s_{i}|i=1,2,\ldots,N\}$
with $N-M$ up-spins and $M$ down-spins,
the Bethe state with $k$ and $\lambda$ 
has the following form:
\begin{align}
\label{eq:Bethe-state}
 |k,\lambda;s\rangle
 =\sum_{\{x_{i}\}}\psi_{k,\lambda}(x;s)
  c_{x_{1},s_{1}}^{\dagger}c_{x_{2},s_{2}}^{\dagger}\cdots 
  c_{x_{N},s_{N}}^{\dagger}|0\rangle.
\end{align}
The coefficients $\psi_{k,\lambda}(x;s)$ 
in \eqref{eq:Bethe-state} are expressed ~\cite{Woynarovich_82JPC} as
\begin{align}
 &\psi_{k,\lambda}(x;s)
  =\sum_{P\in \mathfrak{S}_{N}}\mathrm{sign}(PQ)
   \varphi_{k_{P},\lambda}(s_{Q})
   \exp\Big(\i\sum_{i=1}^{N}k_{P(i)}x_{Q(i)}\Big), \nn\\
 &\varphi_{k_{P},\lambda}(s_{Q})
  =\sum_{R\in\mathfrak{S}_{M}}
   \prod_{\alpha<\beta}
     \frac{\lambda_{R(\alpha)}\!-\!\lambda_{R(\beta)}\!-\!\i U/2}
          {\lambda_{R(\alpha)}\!-\!\lambda_{R(\beta)}}
   \prod_{\gamma=1}^{M}F_{k_{P}}(\lambda_{R(\gamma)},y_{\gamma}), \nn\\
 &F_{k_{P}}(\lambda_{\alpha},y)
  =\frac{1}{\lambda_{\alpha}\!-\!\sin k_{P(y)}\!+\!\i U/4}
   \prod_{j=1}^{y-1}
   \frac{\lambda_{\alpha}\!-\!\sin k_{P(j)}\!-\!\i U/4}
        {\lambda_{\alpha}\!-\!\sin k_{P(j)}\!+\!\i U/4}, \nn
\end{align}
where we have denoted by $Q$ 
one of the shortest elements in the symmetric group 
$\mathfrak{S}_{N}$ on $\{1,2,\ldots,N\}$
such that $1\leq x_{Q(1)}\leq x_{Q(2)}\leq\cdots\leq x_{Q(N)}\leq L$,
and by $y_{\gamma}$ the position of the $\gamma$th down-spin 
in $s_{Q}=\{s_{Q(1)},s_{Q(2)},\ldots,s_{Q(N)}\}$.
The Bethe states~\eqref{eq:Bethe-state} give
eigenstates of the Hamiltonian~\eqref{eq:Hubbard_Ham}
if $\{k_{i},\lambda_{\alpha}\}$ satisfy the following equations:
\begin{align}
\label{eq:Lieb-Wu_eq}
 &\e^{\sqrt{-1}k_{i}L}=
  \prod_{\beta=1}^{M}
  \frac{\lambda_{\beta}-\sin k_{i}-\i U/4}
       {\lambda_{\beta}-\sin k_{i}+\i U/4}, \nn\\
 &\prod_{i=1}^{N}
  \frac{\lambda_{\alpha}-\sin k_{i}-\i U/4}
       {\lambda_{\alpha}-\sin k_{i}+\i U/4}
  =\prod_{\beta(\neq \alpha)}
   \frac{\lambda_{\alpha}-\lambda_{\beta}-\i U/2}
        {\lambda_{\alpha}-\lambda_{\beta}+\i U/2},
\end{align}
which are coupled nonlinear equations called Lieb--Wu equations.
The Lieb--Wu equations have not been solved analytically.
But it predicts some important results on thermodynamic properties
of the system through Takahashi's string 
hypothesis~\cite{Takahashi_72PTP,Takahashi_74PTP,Korepin-Essler%
,Takahashi_book}.

The Bethe states~\eqref{eq:Bethe-state} are not only eigenstates
of the Hamiltonian~\eqref{eq:Hubbard_Ham} but also
those of the translation operator $\tau$
and the higher conserved operators $I_{2}$ and $I_{3}$.
By using the solutions $\{k_{i},\lambda_{\alpha}\}$ 
of the Lieb--Wu equations~\eqref{eq:Lieb-Wu_eq}, 
the eigenvalues of $\tau$ and $\{I_{n}\}$ are written as
\begin{align}
\label{eq:eigenvalues_cop}
 &\tau|k,\lambda;s\rangle=\e^{\i\frac{2\pi}{L}P}|k,\lambda;s\rangle,\quad
  P=\frac{L}{2\pi}\Big(\sum_{i=1}^{N}k_{i}\Big)\,(\mathrm{mod}\, L),\nn\\
 &I_{n}|k,\lambda;s\rangle=E_{n}|k,\lambda;s\rangle,\quad
  (n=1,2,3), \nn\\
 &E_{1}=E=-2\sum_{i=1}^{N}
  \cos k_{i}+\frac{1}{4}U(L\!-\!2N), \nn\\
 &E_{2}=-2\sum_{i=1}^{N}
  \big(\sin(2k_{i})+U\sin k_{i}\big), \nn\\
 &E_{3}=-\sum_{i=1}^{N}
  \Big(
  2\cos(3k_{i})
  +3U\Big(\!\cos(2k_{i})-\frac{1}{2}\Big)+U^{2}\cos k_{i}
  \Big)-\frac{3}{4}UL. 
\end{align}
The $P$ appearing in the eigenvalues of $\tau$ indicates
the total momentum of the system.
(There should be no confusion with the use of $P$ in the coefficients
of the Bethe states where $P$ denotes an element in $\mathfrak{S}_{N}$.)

We immediately find
\begin{align}
\label{eq:Sz-etaz}
 &S_{z}|k,\lambda;s\rangle
  =\frac{1}{2}(N-2M)|k,\lambda;s\rangle,\quad
  \eta_{z}|k,\lambda;s\rangle
  =\frac{1}{2}(L-N)|k,\lambda;s\rangle.
\end{align}
It is shown in~\cite{Essler-Korepin_92aNPB} that
each Bethe state~\eqref{eq:Bethe-state} 
with a regular solution for the Lieb--Wu 
equations~\eqref{eq:Lieb-Wu_eq} corresponds to
the highest weight vector of a highest weight representation
of $\mathfrak{so}(4)$, 
i.e., $S_{+}|k,\lambda;s\rangle=\eta_{+}|k,\lambda;s\rangle=0$.
Then we find
\begin{align}
 \vecvar{S}^{2}|k,\lambda;s\rangle
  =S(S+1)|k,\lambda;s\rangle,\quad
 \vecvar{\eta}^{2}|k,\lambda;s\rangle
  =\eta(\eta+1)|k,\lambda;s\rangle,
\end{align}
with $S=(N-2M)/2$ and $\eta=(L-N)/2$.
Hence, by applying the lowering operators 
$(S_{-})^{n}, (0\leq n\leq N-2M)$ and
$(\eta_{-})^{m}, (0\leq m\leq L-N)$ to 
the Bethe states~\eqref{eq:Bethe-state}, we also have 
eigenstates of the Hubbard Hamiltonian~\eqref{eq:Hubbard_Ham},
\begin{align}
\label{eq:eigenvectors}
  |k,\lambda;s;n,m\rangle\Define
  (S_{-})^{n}(\eta_{-})^{m}|k,\lambda;s\rangle,\quad
\end{align}
which have the same eigenvalues $\{E_{n}\}$ for the operators 
$\{I_{n}\}$ as those of $|k,\lambda;s\rangle$.
The dimension of the representation with the highest weight vector
$|k,\lambda;s\rangle$ is $(N-2M+1)(L-N+1)$.
By using Takahashi's string hypothesis for 
the Bethe states~\eqref{eq:Bethe-state} together with
the $\mathfrak{so}(4)$ symmetry,  
their combinatorial completeness is proved
in~\cite{Essler-Korepin_92bNPB}.

\subsection{Lieb--Wu equations with one down-spin}
\label{sec:LW-eq_one-down}

We try to find regular solutions of 
the Lieb--Wu equations~\eqref{eq:Lieb-Wu_eq}
in the case when the system has only one down-spin
following the discussion in~\cite{Deguchi_00PR}.
In this case, the string hypothesis~\cite{Takahashi_72PTP}
predicts that two types of solutions exist:
one is the solution with only real wavenumbers $\{k_{i}\}$ and
another includes two complex wavenumbers.
First we consider the real wavenumber solutions. 
For $M=1$, the Lieb--Wu equations~\eqref{eq:Lieb-Wu_eq} reduce to
\begin{align}
\label{eq:LW_1-down-spin}
 &\e^{\i k_{i}L}=
  \frac{\lambda-\sin k_{i}-\i U/4}
       {\lambda-\sin k_{i}+\i U/4},\quad(i=1,2,\ldots,N), \nn\\
 &\prod_{i=1}^{N}
  \frac{\lambda-\sin k_{i}-\i U/4}{\lambda-\sin k_{i}+\i U/4}=1. 
\end{align}
These are equivalent to the following equations:
\begin{align}
\label{eq:LW_real-k}
 \sin k_{i}-\lambda=\frac{U}{4}\cot\Big(\frac{k_{i}L}{2}\Big), \quad
 \exp\Big(\i\sum_{i=1}^{N}k_{i}L\Big)=1.
\end{align}
We consider the real solutions 
for the first equation
\begin{align}
\label{eq:LW_real-q}
 \sin q-\lambda=\frac{U}{4}\cot\Big(\frac{qL}{2}\Big).
\end{align}
In the interval $0\leq q<2 \pi$,
its right hand side has $L$ branches 
\begin{align}
\label{eq:branch}
 \frac{2\pi}{L}\Big(\ell-\frac{1}{2}\Big)
 <q<
 \frac{2\pi}{L}\Big(\ell+\frac{1}{2}\Big),\quad
 \ell\in\Big\{\frac{2j-1}{2}\Big|j=1,2,\ldots,L\Big\}.
\end{align}
If we seek a solution $q$ of 
\eqref{eq:LW_real-q} in one of the branches~\eqref{eq:branch}, 
the solution is unique under the condition $UL>8$~\cite{Deguchi_00PR}
and can be written as a function of $\lambda$, i.e.,
$q=q_{\ell}(\lambda)$.
Given a distinct set $\{\ell_{i}|i=1,2,\ldots,N\}\subset
\{\frac{2j-1}{2}|j=1,2,\ldots,L\}$ of the branches,
the second equation in \eqref{eq:LW_real-q} is satisfied
when $\frac{L}{2\pi}\sum_{i=1}^{N}q_{\ell_{i}}(\lambda)\in\mathbb{Z}$.
The behaviour of the solution $q=q_{\ell}(\lambda)$
tells us that
\begin{align}
\label{eq:limit_q-lambda}
  \lim_{\lambda\to\pm\infty}\frac{L}{2\pi}
  \sum_{i=1}^{N}q_{\ell_{i}}(\lambda)
  =\sum_{i=1}^{N}\Big(\ell_{i}\pm\frac{1}{2}\Big).
\end{align}
Hence there exist $N-1$ values of $\lambda$ 
which give the following integer values
for $\frac{L}{2\pi}\sum_{i=1}^{N}q_{\ell_{i}}(\lambda)$:
\[
  m\in
  \Bigg\{\sum_{i=1}^{N}\Big(\ell_{i}-\frac{1}{2}\Big)+j
  \Bigg|j=1,2,\cdots,N-1\Bigg\}.
\]
Note that such $\{\lambda\}$ and integers $\{m\}$ are in
one-to-one correspondence due to $\frac{dq_{\ell}(\lambda)}{d\lambda}>0$.
It is straightforward that 
$\{k_{i}=q_{\ell_{i}}(\lambda),\lambda\}$ characterized by
the indices $\{\ell_{i},m\}$ give 
$\big({L \atop N}\big)(N-1)$
solutions of the equations~\eqref{eq:LW_real-k}.

Next we consider the $k$-$\Lambda$-string solutions.
We assume the forms of solutions as
\[
  k_{i}\in\mathbb{R}/2\pi\mathbb{R},\;(i=1,2,\ldots,N-2),\quad
  k_{N-1}=\zeta\!-\!\i\xi,\quad
  k_{N}=\zeta\!+\!\i\xi,
\] 
where $0\leq \zeta<2\pi$ and $\xi>0$.
Note that $k_{N-1}$ and $k_{N}$ form a complex conjugate pair
which is referred to as $k$-$\Lambda$-two-string.
Then the first equations in \eqref{eq:LW_1-down-spin}
are rewritten as the following equations with real variables:
\begin{subequations}
\label{eq:LW_string-k}
\begin{align}
\label{eq:LW_string-k_1}
 &\sin k_{i}-\lambda
  =\frac{U}{4}\cot\Big(\frac{k_{i}L}{2}\Big),\quad
 (i=1,2,\ldots,N-2), \\
\label{eq:LW_string-k_2}
 &\sin\zeta \cosh\xi-\lambda
  =\frac{U}{4}\frac{\sin(\zeta L)}{\cosh(\xi L)-\cos(\zeta L)}, \\
\label{eq:LW_string-k_3}
 &\cos\zeta \sinh\xi
  =-\frac{U}{4}\frac{\sinh(\xi L)}{\cosh(\xi L)-\cos(\zeta L)}.
\end{align}
\end{subequations}
On the other hand,
the second equation in~\eqref{eq:LW_1-down-spin} 
is equivalent to the following condition:
\begin{align}
\label{eq:LW_k-zeta}
 \sum_{i=1}^{N-2}k_{i}+2\zeta=\frac{2\pi}{L}m,\quad
 (m\in\{0,1,\ldots,NL-1\}).
\end{align}
In the same way as the previous case, 
if we consider a solution of each equation \eqref{eq:LW_string-k_1}
in one of the branches~\eqref{eq:branch},
it can be written as a function of $\lambda$.
Given a distinct set $\{\ell_{i}|i=1,2,\ldots,N-2\}$ of indices 
specifying the branches~\eqref{eq:branch}, 
we express the solutions of~\eqref{eq:LW_string-k_1}
as $k_{i}=q_{\ell_{i}}(\lambda), (i=1,2,\ldots,N-2)$.
Then, from the relation \eqref{eq:LW_k-zeta}, 
the $\zeta$ is also written 
as a function of $\lambda$,
\begin{align}
\label{eq:zeta-lambda}
  \zeta=\zeta(\lambda)\Define\frac{\pi}{L}m
  -\frac{1}{2}\sum_{i=1}^{N-2}q_{\ell_{i}}(\lambda),
\end{align}
for fixed $\{\ell_{i}\}$ and $m$.
For an illustration, we consider \eqref{eq:LW_string-k_2} and
\eqref{eq:LW_string-k_3} in the case $N=2$.
Since $\zeta$ does not depend on $\lambda$ in the case,
the equations \eqref{eq:LW_string-k_2} and
\eqref{eq:LW_string-k_3} decouple into
\begin{align}
 \label{eq:LW_string-k_N=2}
 &\lambda=\sin\Big(\frac{\pi}{L}m\Big)\cosh\xi, \quad
 \sinh\xi=-\frac{U}{4\cos(\frac{\pi}{L}m)}f^{(2)}(\xi),
\end{align}
where
\[
f^{(2)}(\xi)\Define\frac{\sinh(\xi L)}{\cosh(\xi L)-(-1)^{m}}=
  \begin{cases}
   \tanh(\xi L/2) &\text{for } m \text{ odd}, \\
   \coth(\xi L/2) &\text{for } m \text{ even}. \\
  \end{cases}
\]
One finds from graphical discussion~\cite{Deguchi_00PR}
that, if the condition $UL>8$ is satisfied,
the second equation determines an unique $\xi(>0)$ for 
$\frac{\pi}{2}<\frac{\pi}{L}m<\frac{3\pi}{2}$.
We denote it as $\xi_{m}^{(2)}, 
(m\in\{\frac{L}{2}+j|j=1,2,\cdots,L-1\})$.
The first equation in \eqref{eq:LW_string-k_N=2}
immediately gives $\lambda$ with the $\xi_{m}^{(2)}$.
Let us consider the case $N>2$.
By inserting \eqref{eq:zeta-lambda} into \eqref{eq:LW_string-k_3},
we have
\begin{align}
\label{eq:LW_xi-q}
  \sinh\xi=-\frac{U}{4\cos(\zeta(\lambda))}f(\xi),\quad
  f(\xi)\Define
  \frac{\sinh(\xi L)}{\cosh(\xi L)
  -(-)^{m}\cos\big(\frac{L}{2}\sum_{i}q_{\ell_{i}}(\lambda)\big)}. 
\end{align}
Since $f(\xi)>0$ for $\xi>0$ in the similar to the case $N=2$,
this determines an unique $\xi$ as a function of $\lambda$ 
if and only if 
$\frac{\pi}{2}<\zeta(\lambda)=\frac{\pi}{L}m
 -\frac{1}{2}\sum_{i}q_{\ell_{i}}(\lambda)
 <\frac{3\pi}{2}$.
By using \eqref{eq:limit_q-lambda},
it is sufficient to have an unique solution for \eqref{eq:LW_xi-q}
that the integer $m$ satisfies
\[
  \sum_{i=1}^{N-2}\Big(\ell_{i}+\frac{1}{2}\Big)+\frac{L}{2}
  <m<
  \sum_{i=1}^{N-2}\Big(\ell_{i}-\frac{1}{2}\Big)+\frac{3L}{2},
\]
that is, 
\begin{align}
\label{eq:string_m}
  m\in\Bigg\{\sum_{i=1}^{N-2}\Big(\ell_{i}+\frac{1}{2}\Big)
             +\frac{L}{2}+j\Bigg|j=1,2,\ldots,L-N+1\Bigg\}.
\end{align}
Note that $\lim_{\lambda\to\pm\infty}\xi(\lambda)
=\xi_{m-\sum_{i}(\ell_{i}\pm\frac{1}{2})}^{(2)}$ which
is well-defined for the above $m$.
We expect that,
for the $L-N+1$ values of $m$ allowed in \eqref{eq:string_m},
the equation \eqref{eq:LW_string-k_2} with 
$\xi(\lambda)$ and $\zeta(\lambda)$ 
\begin{align}
\label{eq:LW_string-k_3_1}
 \lambda
 &=\sin\big({\textstyle
 \frac{\pi}{L}m\!-\!\frac{1}{2}\sum_{i}q_{\ell_{i}}(\lambda)}\big)
  \cosh\big(\xi(\lambda)\big)
 -\frac{U}{4}\frac{\sin(\frac{L}{2}\sum_{i}q_{\ell_{i}}(\lambda))}
 {\cos\big(\frac{L}{2}\sum_{i}q_{\ell_{i}}(\lambda)\big)
  -(-)^{m}\cosh\big(\xi(\lambda) L\big)} \nn\\
 &\definE g\big(q_{\ell_{i}}(\lambda),\xi(\lambda)\big),
\end{align}
determines $\lambda$.
In fact, since $q_{\ell_{i}}(\lambda)$ and $\xi(\lambda)$
are continuous functions with respect to $\lambda$ and
\[
 \lim_{\lambda\to\pm\infty}
 g\big(q_{\ell_{i}}(\lambda),\xi(\lambda)\big)
 =g\Big({\textstyle
  \frac{2\pi}{L}(\ell_{i}\pm\frac{1}{2}),
  \xi_{m-\sum_{i}(\ell_{i}\pm\frac{1}{2})}^{(2)}}\Big),
\]
$g$ is a continuous and finite function 
with respect to $\lambda$.
Hence there exists a solution $\lambda$ 
in \eqref{eq:LW_string-k_3_1}.
As a consequence we have $({L \atop N-2})(L-N+1)$ solutions 
corresponding to the indices $\{\ell_{i},m\}$.

Let us see the relation between the string 
hypothesis~\cite{Takahashi_72PTP} and our results.
Let $\{\ell_{i}\}$ reduce modulo $L$ into the interval
$[-\frac{L}{2},\frac{L}{2})$ and express them as $\{I^{(i)}\}$.
Putting $J=m-\sum_{i=1}^{N}\ell_{i}$ for the real wavenumber solutions 
and $J^{\prime}=L-m+\sum_{i=1}^{N-2}\ell_{i}$ for the 
$k$-$\Lambda$-two-string solution, we have
\begin{align}
\label{eq:IJJ}
 &I^{(i)}\in\mathbb{Z}+\frac{1}{2},\quad 
  -\frac{L}{2}<I^{(i)}\leq\frac{L}{2}, \nn\\
 &J\in\mathbb{Z}+\frac{N}{2},\quad |J|\leq\frac{1}{2}(N-2), \quad
  J^{\prime}\in\mathbb{Z}+\frac{N}{2},\quad
  |J^{\prime}|\leq\frac{1}{2}(L-N).
\end{align}
One sees that the indices $\{I^{(i)};J;J^{\prime}\}$ 
characterizing the regular solutions of the Lieb--Wu equations
with $M=1$ are nothing but those appearing 
in the string hypothesis~\cite{Takahashi_72PTP}.
Thus we have shown for the Lieb--Wu equations~\eqref{eq:Lieb-Wu_eq}
with $M=1$ that there exist the same number of solutions
as those predicted by the string hypothesis.
In the next section, we numerically calculate 
the solutions $\{k_{i},\lambda|i=1,2,\ldots,N\}$ and
$\{k_{i},\zeta,\xi,\lambda|i=1,2,\ldots,N-2\}$ for $L=6$.

\section{Energy level crossings}
\label{sec:crossings}

Let us review von Neumann--Wigner's discussion 
on the spectra of quantum systems.
We assume that a Hamiltonian is described by 
a finite-dimensional real symmetric 
matrix whose elements are regarded as random independent parameters.
If the system has no symmetry, 
we call its spectra a ``pure sequence''.
If there exist some symmetries in the system, its spectra 
is given by a superposition of pure sequences,
which we call a ``mixed sequence''.
Von Neumann--Wigner's theorem reads as follows:
one must adjust two parameters to bring two of the eigenvalues 
belonging to a pure sequence into coincidence 
while in a mixed sequence one obtain a degeneracy 
by varying only one parameter~\cite{Neumann-Wigner_29PZ,Reichl}.
Hence the degeneracies in pure sequences are very unlikely to be found
if we choose the values of parameters in an arbitrary manner.
Such degeneracies in pure sequences are referred to as
accidental degeneracies.

Applying the above discussion, we study the Hubbard model.
In varying the parameter $U$,
the Hamiltonian~\eqref{eq:Hubbard_Ham} gives a flow 
in the above space of parameters. 
As we have mentioned in the previous section,
the system has several symmetries.
Since the operators 
$\{\tau,\vecvar{S}^{2},S_{z},\vecvar{\eta}^{2},\eta_{z}\}$
are mutually commutative, they can be simultaneously diagonalized 
by an orthogonal transformation.
Through the same orthogonal transformation,
the Hamiltonian matrix breaks up into 
diagonal blocks corresponding to the common eigenspaces
of $\{\tau,\vecvar{S}^{2},S_{z},\vecvar{\eta}^{2},\eta_{z}\}$.
Notice that the common eigenspaces are characterized by 
the set of quantum numbers $\{N,M,P,S,\eta\}$.
Energy eigenvalues from the blocks with different quantum numbers
may degenerate due to translation and $SO(4)$ symmetries.
But all the blocks do not give a pure sequence;
for example, the blocks with $L=N=2M$ have particle-hole and
spin reversal symmetries.
However, after considering all the known $U$-independent symmetries,
the spectra also have many degeneracies at special values of $U$, 
i.e., level crossings in the spectral flows along the parameter 
$U$~\cite{Heilmann-Lieb_71}. 
Thus the flow determined by the Hamiltonian~\eqref{eq:Hubbard_Ham}
runs through several special values of parameters 
that give accidental degeneracies.

Let us discuss how to numerically determine level crossings, 
in particular, for the case of accidental degeneracies. 
One notices that the numerical diagonalization of the Hamiltonian 
is not enough to find crossings of energy spectral flows 
since apparent crossings may be just close approaches of 
two eigenvalues~\cite{Heilmann-Lieb_71}.
To verify the existence of genuine level crossings,
we must investigate the behaviour of spectral flows for each eigenstate.
The Bethe ansatz method is a very effective tool to establish this.
Here we restrict our investigation to the systems with one down-spin where
we have shown the existence of solutions for the Lieb--Wu 
equations~\eqref{eq:Lieb-Wu_eq}
in Subsection~\ref{sec:LW-eq_one-down}.
We have seen in Subsection~\ref{sec:symmetries}
that the Hubbard model has the dynamical symmetries
in addition to the $U$-independent symmetries.
The recent paper~\cite{Shastry_02JPA} pointed out that 
the degenerate eigenstates 
at the accidental degeneracies observed in \cite{Heilmann-Lieb_71}
can be classified by the eigenvalues of the higher conserved operators. 
We verify their assertion at the level of eigenstates in the  
sector of one down-spin. 
Here we note that the degeneracies discussed in 
\cite{Heilmann-Lieb_71,Bondeson-Soos_79JCP,Shastry_02JPA} are 
in the half-filled Hubbard model with zero magnetization,
that is, in the sector of three down-spins.
The strategy in the present paper is given by the following:
\begin{itemize}
\item[i)] We give a matrix representation of the Hamiltonian
in a certain common eigenspace of the operators
$\{\tau,\vecvar{S}^{2},S_{z},\vecvar{\eta}^{2},\eta_{z}\}$,
which is referred to as desymmetrization of
the corresponding symmetries.
We numerically give its eigenvalues from the direct 
diagonalization~\cite{Heilmann-Lieb_71,Shastry_02JPA} 
for several values of $U$.
\item[ii)] We give the numerical solutions of the Lieb--Wu 
equations~\eqref{eq:Lieb-Wu_eq} in several values of $U$.
{}From the correspondence between energy spectral flows
and the Bethe states, we verify whether or not
genuine energy level crossings exist.
\item[iii)] 
We also diagonalize other conserved operators 
$\{I_{2},I_{3}\}$ and see the correspondence between 
their spectral flows and the Bethe states. 
We analyse the structure of their degeneracies.
\end{itemize}
We discuss only the systems containing 
two or three up-spins and one down-spin,
which do not have particle-hole and spin reversal symmetries.
In spite of such restriction, 
we have some non-trivial results on degeneracies.

\subsection{Twofold permanent degeneracies}
\label{sec:twofold}

The translation and the $SO(4)$ symmetries 
produce $U$-independent degeneracies in energy spectral flows.  
We call $U$-independent degeneracies permanent degeneracies. 
Furthermore, after the desymmetrization of these 
$U$-independent symmetries,
we often observe another twofold permanent degeneracies 
associated with a reflection symmetry of the lattice~\cite{Shastry_02JPA}. 
In fact, we can explain them in terms of the Bethe ansatz wavefunction. 
Here we note that, due to the relation $\tau\sigma=\sigma\tau^{-1}$, 
the reflection operator $\sigma$ acts only on the eigenspaces of $\tau$ 
with the eigenvalue $1$ or $-1$, 
that is, the subspaces of the Fock space with 
the total momentum $P=0$ or $\frac{L}{2}$.

Let us investigate the twofold permanent degeneracies due to the reflection 
operator $\sigma$ in the framework of the Bethe ansatz method.
Even in the subspaces with $P=0$ or $\frac{L}{2}$, 
the Bethe states do not always diagonalize
the operator $\sigma$.
Indeed it is easy to verify that,
if we apply the operator $\sigma$ to one of the eigenvectors
$|k,\lambda;s;m,n\rangle$, then its total momentum and 
eigenvalues of $\{I_{2n}\}$ are negated and those of 
$\{\vecvar{S}^{2},S_{z},\vecvar{\eta}^{2},\eta_{z},I_{2n-1}\}$ 
do not change since
\begin{align}
\label{eq:sigma-cop}
  \sigma\tau=\tau^{-1}\sigma,\quad
  \sigma I_{2n}+I_{2n}\sigma=0,\quad
  \sigma I_{2n-1}=I_{2n-1}\sigma,\quad
  [\sigma,S_{z,\pm}]=[\sigma,\eta_{z,\pm}]=0.
\end{align}
These facts are verified more directly 
from the following relation:
\begin{align}
\label{eq:sigma-to-state}
 \sigma|k,\lambda;s;n,m\rangle
 =(-)^{M}|-k,-\lambda;s;n,m\rangle,
\end{align}
where $-k=\{-k_{i}|i=1,2,\ldots,N\}$ and
$-\lambda=\{-\lambda_{\alpha}|\alpha=1,2,\ldots,M\}$.
Note that, if $\{k_{i},\lambda_{\alpha}\}$ is a solution
for the Lieb--Wu equations~\eqref{eq:Lieb-Wu_eq},
then so is $\{-k_{i},-\lambda_{\alpha}\}$.  
The relation~\eqref{eq:sigma-to-state} means that,
if the set $\{k_{i};\lambda_{\alpha}\}$ does not coincide
with $\{-k_{i};-\lambda_{\alpha}\}$ as a set, we have
a twofold permanent degeneracy in energy spectral flows.
On the other hand, if 
$\{k_{i};\lambda_{\alpha}\}=\{-k_{i};-\lambda_{\alpha}\}$ as a set,
then the eigenstates $|k,\lambda;s;m,n\rangle$ have 
the eigenvalue $E_{2}=0$ for the second conserved operator $I_{2}$,
which follows from the formulas in \eqref{eq:eigenvalues_cop}.
It should be noted that, 
to see the existence of such twofold permanent degeneracies,
we must solve the Lieb--Wu equations~\eqref{eq:Lieb-Wu_eq}.

\subsection{Spectral flows: $L=6, N=3, M=1$}
\label{sec:631}

\begin{figure}[p]
\begin{center}
\includegraphics[width=100mm,clip]{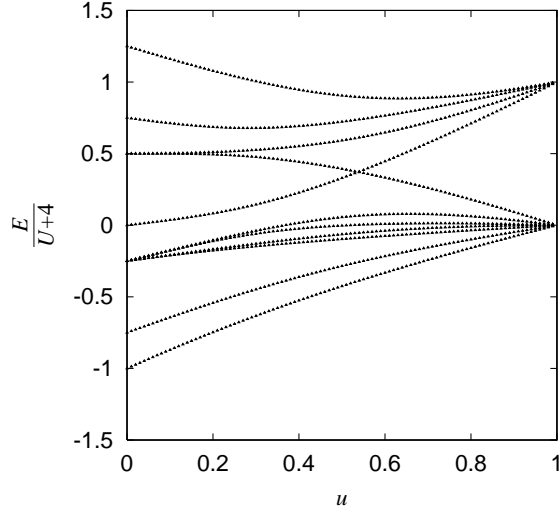}
\end{center}
\caption{Spectral flows $E$ from direct diagonalization
for $L=6, N=3, M=1$, $P=2, S=\frac{1}{2}$ and $\eta=\frac{3}{2}$.}
\label{fig:E1-631}
\end{figure}

We now study the system with two up-spins and one down-spin
on benzene ($L=6, N=3, M=1$).
Note that $S_{z}=\frac{1}{2}$ and $\eta_{z}=\frac{3}{2}$ in this case.
We consider the subspace characterized by the set of quantum numbers
$\{P=2,S=\frac{1}{2},\eta=\frac{3}{2}\}$.
There is no more $U$-independent symmetry in the subspace, and
we shall actually find one of the simplest nontrivial energy level 
crossings there.
We have a matrix representation of the Hamiltonian of $11$ dimension.
We put $u\Define U/(U+4)$ and numerically diagonalize 
the Hamiltonian matrix for several values of $u$.
Figure \ref{fig:E1-631} shows the energy spectral flows for $0<u<1$.
Here the vertical line in the figure indicates 
the energy $E$ divided by $U+4$.
We confirm from the numerical data
that there is no permanent degeneracy.
We observe a close approach of two energy levels in $0.5<u<0.6$,
which seems to be an energy level crossing.

Our first purpose is to show that 
a genuine energy level crossing has been found in Figure \ref{fig:E1-631}.
In the case of triangular quantum billiards~\cite{Berry-Wilkinson_84PRSL}, 
topological properties of their eigenstates were studied
to verify the existence of energy level crossings.
Here we employ the Bethe ansatz method to verify energy level crossings
in the level of eigenstates.
It is clear that, since $S=S_{z}=\frac{1}{2}$ and 
$\eta=\eta_{z}=\frac{3}{2}$, all the eigenstates 
in the subspace are the Bethe states
characterized by the indices $\{I^{(i)};J;J^{\prime}\}$
satisfying \eqref{eq:IJJ}.
By using the procedure in Subsection \ref{sec:LW-eq_one-down},
we numerically give real wavenumber solutions with
$P=(\sum_{i=1}^{3}I^{(i)}+J)\, (\mathrm{mod}\, 6)
=2\, (\mathrm{mod}\, 6)$
and $k$-$\Lambda$-two-string solutions with
$P=(I^{(1)}-J^{\prime})\, (\mathrm{mod}\, 6)=2\, (\mathrm{mod}\, 6)$
to calculate the corresponding energy eigenvalues.
The correspondence between energy eigenvalues and the Bethe states
at $u=0.3$ and $0.8$ is displayed on Table 1.
Here we deal with the Lieb--Wu equations only
in the case when the condition $UL>8$ is satisfied.
One sees that the results for $u\to 1, (U\to\infty)$ read as
\[
  \lim_{U\to\infty}\frac{E}{U+4}=
  \begin{cases}
  0 & \text{for real wavenumber solutions,}\\
  1 & \text{for $k$-$\Lambda$-two-string solutions,}\\
  \end{cases}
\]
which agrees with those conjectured by the string hypothesis.
We remark that the energy eigenvalues obtained
by the solutions of the Lieb--Wu equations coincide with 
those obtained by the direct diagonalization of the Hamiltonian
matrix within an error of $O(10^{-15})$.
Thus, by combining the results in Figure \ref{fig:E1-631} 
with those on Table \ref{tab:T1-631},
we obtain the behaviour of energy spectral flows for each eigenstate.
We conclude that, in $0.5<u<0.6$, there exists 
an energy level crossing between two Bethe states indexed by
$\{-\frac{5}{2},\frac{3}{2},\frac{5}{2};\frac{1}{2};\}$ and
$\{\frac{1}{2};;-\frac{3}{2}\}$.

\begin{figure}[p]
\begin{center}
\includegraphics[width=100mm,clip]{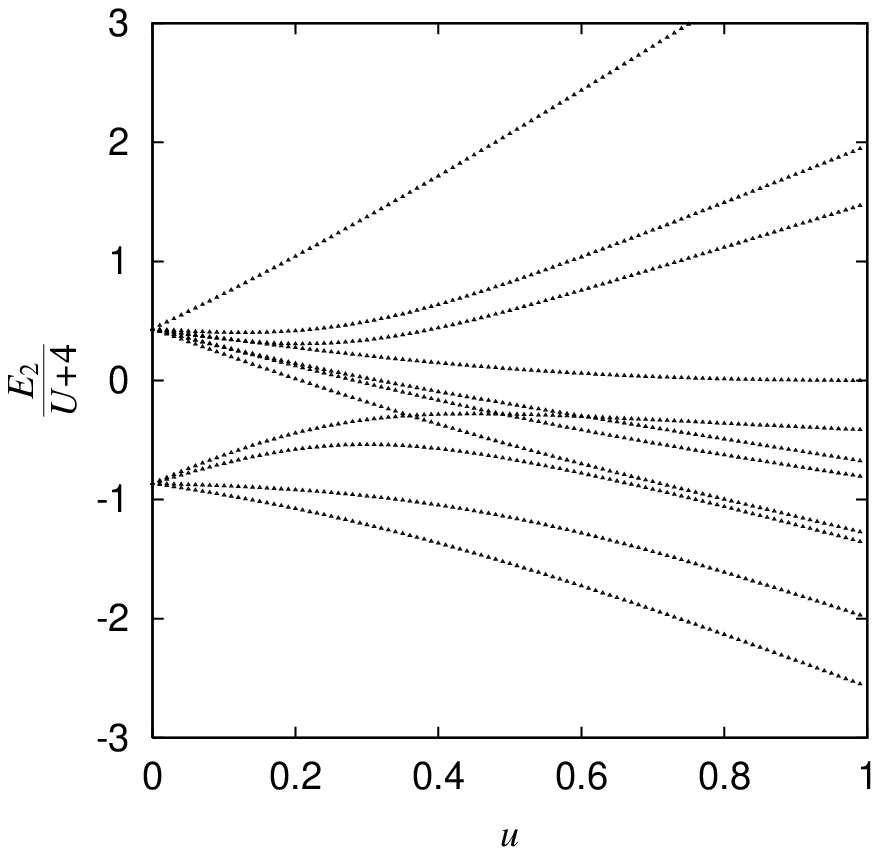}
\end{center}
\caption{Spectral flows $E_{2}$ from direct diagonalization 
for $L=6, N=3, M=1$, $P=2, S=\frac{1}{2}$ and $\eta=\frac{3}{2}$.}
\label{fig:E2-631}
\end{figure}

Next we show that, if we take into account dynamical symmetries,
each one-dimensional component of the degenerate eigenstates 
at the energy level crossing point 
can be distinguished from the other degenerate eigenstates. 
In the similar way to the energy eigenvalues,
the eigenvalues  of the second conserved operator $I_{2}$ are
obtained by both their direct diagonalization
and the solutions of the Lieb--Wu equations,
which are displayed in Figure \ref{fig:E2-631} 
and Table \ref{tab:T2-631}
(the vertical line in the figure also indicates
the eigenvalue $E_{2}$ divided by $U+4$).
Note that there is no permanent degeneracy.
We observe that the spectral flows indexed by
$\{-\frac{5}{2},\frac{3}{2},\frac{5}{2};\frac{1}{2};\}$ and
$\{\frac{1}{2};;-\frac{3}{2}\}$ in Figure~\ref{fig:E2-631} 
never have crossings.
Hence the Bethe states indexed by
$\{-\frac{5}{2},\frac{3}{2},\frac{5}{2};\frac{1}{2};\}$ and
$\{\frac{1}{2};;-\frac{3}{2}\}$ belong to different eigenspaces
of $I_{2}$, that is, the two Bethe states have 
a different dynamical symmetry. 
Thus all the common eigenspaces of the operators 
$\{H,I_{2}\}$ with the set of quantum numbers 
$\{P=2,S=S_{z}=\frac{1}{2},\eta=\eta_{z}=\frac{3}{2}\}$ are one-dimensional.

\begin{table}[p]
\begin{center}
\def\arraystretch{1.5}
\begin{tabular}{|l|l|l|l|}
\hline
 & \multicolumn{3}{|c|}{$E/(U+4)$} \\
\hline
 $\{I^{(i)};J;J^{\prime}\}$
 & \multicolumn{1}{|c|}{$u=0.3$} 
 & \multicolumn{1}{|c|}{$u=0.8$}
 & \multicolumn{1}{|c|}{$u=1.0$} \\
\hline
 $\{\frac{5}{2};;\frac{1}{2}\}$
 & $\;\;1.005067890492477$
 & $\;\;0.9123001584634441$
 & $\;\;1$ \\
\hline
 $\{-\frac{5}{2};;\frac{3}{2}\}$
 & $\;\;0.6786192438740763$
 & $\;\;0.8733443175466562$
 & $\;\;1$ \\
\hline
 $\{\frac{3}{2};;-\frac{1}{2}\}$
 & $\;\;0.5258327541688925$
 & $\;\;0.8031341881381975$
 & $\;\;1$ \\
\hline
 $\{-\frac{5}{2},\frac{3}{2},\frac{5}{2};\frac{1}{2};\}$
 & $\;\;0.4763789806072449$
 & $\;\;0.1794157938506165$
 & $\;\;0$ \\
\hline
 $\{\frac{1}{2};;-\frac{3}{2}\}$
 & $\;\;0.144465968291432$
 & $\;\;0.7105015978852069$
 & $\;\;1$ \\
\hline
 $\{-\frac{3}{2},\frac{3}{2},\frac{5}{2};-\frac{1}{2};\}$
 & $-0.03872530249182613$
 & $\;\;0.06204279614552415$
 & $\;\;0$ \\
\hline
 $\{-\frac{5}{2},-\frac{3}{2},\frac{1}{2};-\frac{1}{2};\}$
 & $-0.05950271399595123$
 & $\;\;0.01208774945632921$
 & $\;\;0$ \\
\hline
 $\{-\frac{3}{2},\frac{1}{2},\frac{5}{2};\frac{1}{2};\}$
 & $-0.1292149141454342$
 & $-0.008632714879121293$
 & $\;\;0$ \\
\hline
 $\{-\frac{5}{2},-\frac{3}{2},-\frac{1}{2};\frac{1}{2};\}$
 & $-0.1472919189221319$
 & $-0.03595459977202529$
 & $\;\;0$ \\
\hline
 $\{-\frac{1}{2},\frac{1}{2},\frac{5}{2};-\frac{1}{2};\}$
 & $-0.447930438860179$
 & $-0.0995848490575883$
 & $\;\;0$ \\
\hline
 $\{-\frac{1}{2},\frac{1}{2},\frac{3}{2};\frac{1}{2};\}$
 & $-0.6326995490186007$
 & $-0.1586544377772396$
 & $\;\;0$ \\
\hline
\end{tabular}
\def\arraystretch{1}
\end{center}
\caption{$E$ obtained by the Bethe ansatz method.}
\label{tab:T1-631}
\end{table}

\begin{table}[p]
\begin{center}
\def\arraystretch{1.5}
\begin{tabular}{|l|l|l|}
\hline
 & \multicolumn{2}{|c|}{$E_{2}/(U+4)$} \\
\hline
 $\{I^{(i)};J;J^{\prime}\}$
 & \multicolumn{1}{|c|}{$u=0.3$} 
 & \multicolumn{1}{|c|}{$u=0.8$}\\
\hline
 $\{-\frac{5}{2},-\frac{3}{2},-\frac{1}{2};\frac{1}{2};\}$
 & $\;\;1.373365167833861$
 & $\;\;3.179687928057096$ \\
 $\{-\frac{5}{2},\frac{3}{2},\frac{1}{2};-\frac{1}{2};\}$
 & $\;\;0.4945223061494427$
 & $\;\;1.493828870394599$ \\
 $\{-\frac{5}{2};;\frac{3}{2}\}$
 & $\;\;0.3397777976721251$
 & $\;\;1.117455777165958$ \\
 $\{-\frac{3}{2},\frac{1}{2},\frac{5}{2};\frac{1}{2};\}$
 & $\;\;0.2072047205613698$
 & $\;\;0.01309975997192909$ \\
 $\{-\frac{1}{2},\frac{1}{2},\frac{5}{2};-\frac{1}{2};\}$
 & $\;\;0.01863957159436896$
 & $-0.4919569001978466$ \\
 $\{\frac{5}{2};;\frac{1}{2}\}$
 & $-0.02590156370174922$
 & $-0.6254191358761778$ \\
 $\{-\frac{3}{2},\frac{3}{2},\frac{5}{2};-\frac{1}{2};\}$
 & $-0.1828197100832381$
 & $-0.9968145682615939$ \\
 $\{\frac{1}{2};;-\frac{3}{2}\}$
 & $-0.3289755229197938$
 & $-0.3593561303568662$ \\
 $\{-\frac{5}{2},\frac{3}{2},\frac{5}{2};\frac{1}{2};\}$
 & $-0.5374256073807519$
 & $-1.059175160002318$ \\
 $\{\frac{3}{2};;-\frac{1}{2}\}$
 & $-0.9702596151665846$
 & $-1.610420144313048$ \\
 $\{-\frac{1}{2},\frac{1}{2},\frac{3}{2};\frac{1}{2};\}$
 & $-1.210851678154266$
 & $-2.13317348301527$ \\
\hline
\end{tabular}
\def\arraystretch{1}
\end{center}
\caption{$E_{2}$ obtained by the Bethe ansatz method.}
\label{tab:T2-631}
\end{table}

\begin{figure}[t]
\quad
\begin{center}
\includegraphics[width=120mm,clip]{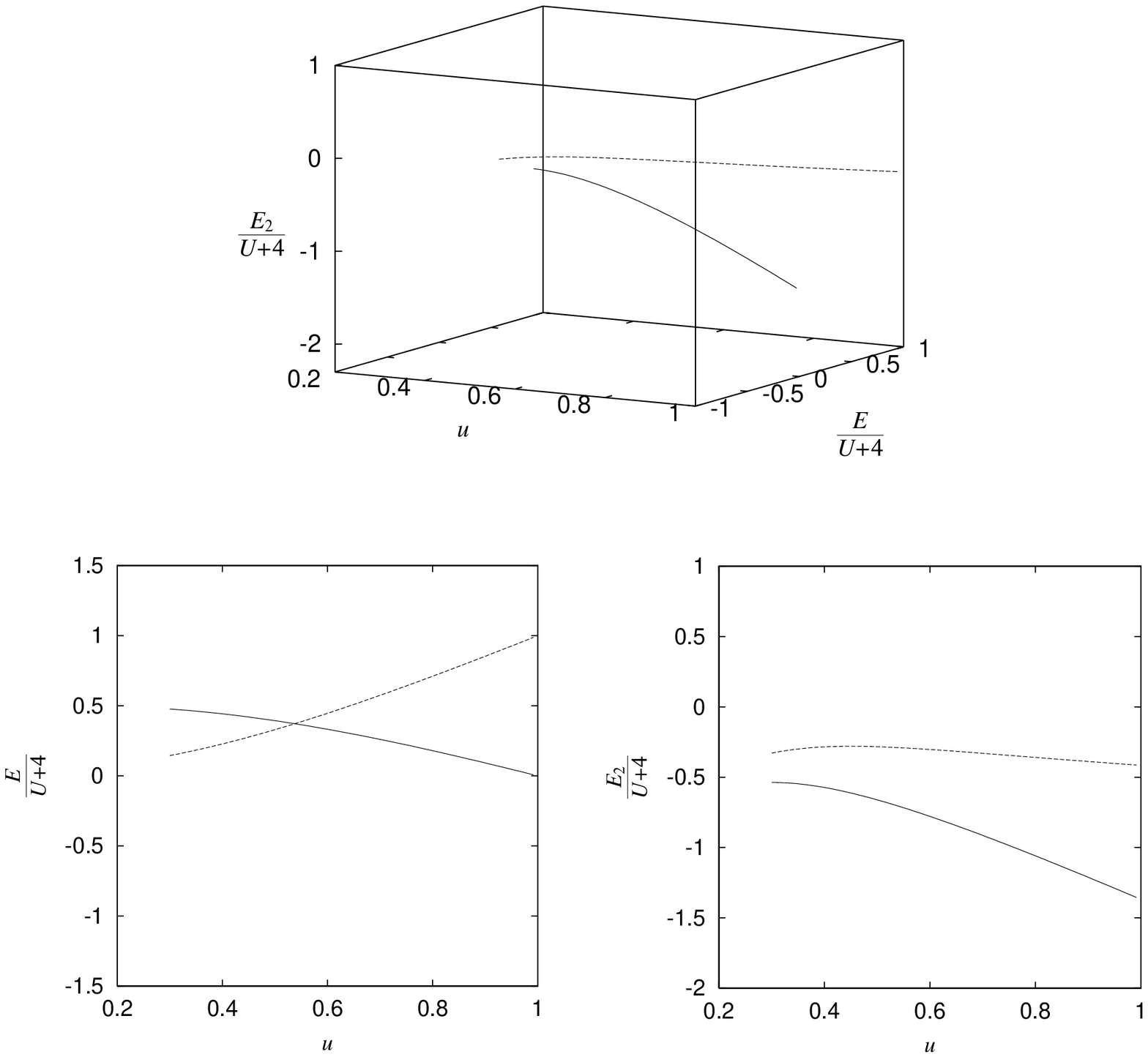}
\end{center}
\caption{Spectral flow sets $(E,E_{2})$ characterized by 
$\{-\frac{5}{2},\frac{3}{2},\frac{5}{2};\frac{1}{2};\}$ 
and $\{\frac{1}{2};;-\frac{3}{2}\}$ for the system with 
$L=6, N=3, M=1$, $P=2, S=\frac{1}{2}$ and $\eta=\frac{3}{2}$,
and their projections onto the $U$-$E$ and $U$-$E_{2}$ planes.}
\label{fig:3d}
\end{figure}

The fact that the common eigenspaces of $\{H,I_{2}\}$
are one-dimensional becomes clearer by investigating
the spectral flows of the transfer matrix, i.e.,
the spectral flow sets of conserved operators.
Figure~\ref{fig:3d} shows that, if we consider the eigenvalue sets 
$\{E,E_{2}\}$ for two of the eigenvectors along the parameter $U$, 
there is no crossing 
while their projection onto the $U$-$E$ plane has a crossing.
One notices that the Bethe ansatz method plays an essential role 
in getting such picture.
It seems to be rare that 
the eigenvalues of the transfer matrix simultaneously degenerate.
We comment that the XXZ Heisenberg spin chain
at root of unity has such degeneracies 
in the transfer matrix~\cite{Deguchi-Fabricius-McCoy_01JSP}, where 
one sees symmetries defined only at the special values of parameters.

\subsection{Spectral flows: $L=6, N=4, M=1$}
\label{sec:641}

\begin{figure}[p]
\begin{center}
\includegraphics[width=100mm,clip]{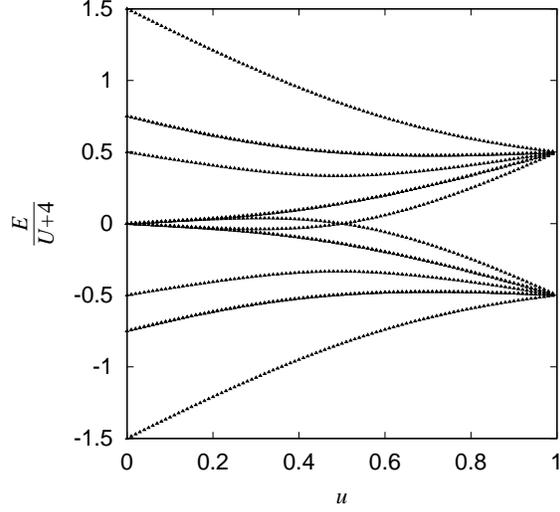}
\end{center}
\caption{Spectral flows $E$ from direct diagonalization
for $L=6, N=4, M=1$, $P=0$ and $S=\eta=1$.}
\label{fig:E1-641}
\end{figure}

We give an example where the eigenvalues $E$ and $E_{2}$
degenerate at the same value of $U$.
We consider the system with three up-spins and one down-spin
on benzene ($L=6, N=4, M=1$).
In this case, we have $S_{z}=\eta_{z}=1$.
We investigate the degeneracies of energy spectral flows 
in the subspace characterized by the set of
quantum numbers $\{P=0,S=\eta=1\}$;
hence we can expect all the eigenstates therein to be the Bethe states.
The subspace is 14 dimensions.
Figure \ref{fig:E1-641} shows the energy spectral flows obtained 
by the numerical diagonalization of the Hamiltonian matrix for $0<u<1$.
Also listed on Table \ref{tab:T1-641} are the energy eigenvalues given by 
numerical solutions for the Lieb--Wu equations at $u=0.4$ and $0.8$;
we show the numerical solutions in Appendix~\ref{sec:numerical}.
We observe as $u\to 1, (U\to\infty)$ that
\[
  \lim_{U\to\infty}\frac{E}{U+4}=
  \begin{cases}
  -\frac{1}{2} & \text{for real wavenumber solutions,}\\
  \frac{1}{2} & \text{for $k$-$\Lambda$-two-string solutions,}\\
  \end{cases}
\]
which are also derived from 
the string hypothesis~\cite{Takahashi_72PTP}.
Here we note that the results from the two different methods above
coincide within an error of $O(10^{-15})$, which give an evidence for
the validity of the Bethe ansatz method.
{}From Figure \ref{fig:E1-641} and Table \ref{tab:T1-641},
we see the correspondence between the energy spectral flows
and the indices $\{I^{(i)};J;J^{\prime}\}$ \eqref{eq:IJJ}
characterizing the Bethe states.
The correspondence shows that, if
$\{I^{(i)};J;J^{\prime}\}\neq \{-I^{(i)};-J;-J^{\prime}\}$, i.e.,
$\{k_{i};\lambda\}\neq\{-k_{i};-\lambda\}$ as a set, 
two Bethe states characterized by these indices are in 
a permanent degeneracy.
In fact, among the 14 Bethe states, 8 of them are 
in the twofold permanent degeneracies associated with 
the reflection symmetry.
We find that the Bethe states indexed by 
$\{-\frac{5}{2},-\frac{3}{2},\frac{3}{2},\frac{5}{2};0;\}$ 
and $\{-\frac{1}{2},\frac{1}{2};;0\}$
has an energy level crossing at $u=0.5, (U=4)$.
Although the existence of this energy level crossing 
is also verified from the characteristic equation
of the Hamiltonian matrix,
we need the Bethe ansatz method
to clarify which Bethe states have the crossing.

\begin{figure}[p]
\begin{center}
\includegraphics[width=100mm,clip]{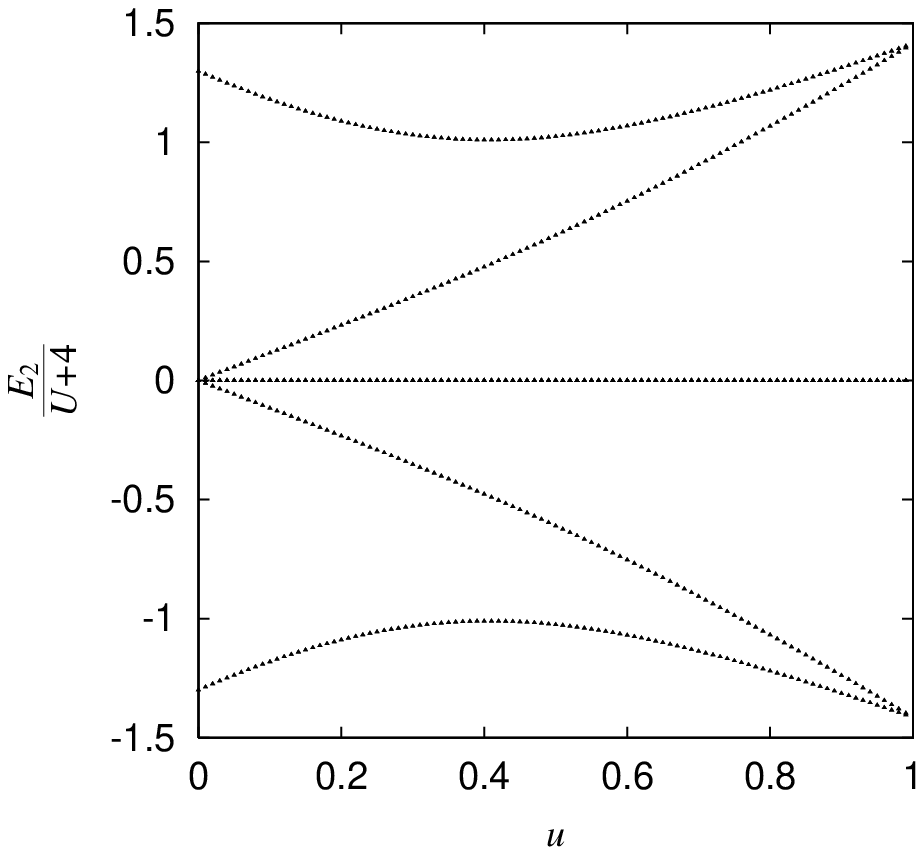}
\end{center}
\caption{Spectral flows $E_{2}$ from direct diagonalization
for $L=6, N=4, M=1$, $P=0$ and $S=\eta=1$.}
\label{fig:E2-641}
\end{figure}

We discuss whether or not each of the above degenerate eigenstates 
have distinct eigenvalues of the higher conserved operators.
In the similar way to the previous case,
we consider the correspondence between the spectral flows of
the second conserved operator $I_{2}$ and the Bethe states,
which is displayed in Figure \ref{fig:E2-641} and Table \ref{tab:T2-641}.
As is described in Subsection~\ref{sec:twofold}, if we express
the eigenvalue of $I_{2}$ for the eigenstate $|k,\lambda;s\rangle$
as $E_{2}$, that for the state $\sigma|k,\lambda;s\rangle$ is $-E_{2}$.
In particular, if 
$\{I^{(i)};J;J^{\prime}\}=\{-I^{(i)};-J;-J^{\prime}\}$, i.e., 
$\{k_{i};\lambda\}=\{-k_{i};-\lambda\}$, the state has $E_{2}=0$.
Figure \ref{fig:E2-641} and Table \ref{tab:T2-641} indeed show that
two eigenstates whose energy spectral flows are in a permanent
degeneracy belong to the different eigenspaces of $I_{2}$ and
there are $6(=14-8)$ states in the eigenspace of $I_{2}$ with $E_{2}=0$.
Thus we may conclude that the dynamical symmetry $I_{2}$ ``accounts for'' 
all the twofold permanent degeneracies in the energy spectral flows 
which are produced by the reflection symmetry.
It is a further remarkable fact that both the Bethe states indexed
by $\{-\frac{5}{2},-\frac{3}{2},\frac{3}{2},\frac{5}{2};0;\}$ 
and $\{-\frac{1}{2},\frac{1}{2};;0\}$, which has 
an energy level crossing, belong to the same eigenspaces of $I_{2}$
with $E_{2}=0$.
Hence, at $u=0.5, (U=4)$,
the two Bethe states can not be distinguished from their eigenvalues of
$\{\tau,\vecvar{S}^{2},S_{z},\vecvar{\eta}^{2},\eta_{z},H,I_{2}\}$.

\begin{figure}[p]
\begin{center}
\includegraphics[width=100mm,clip]{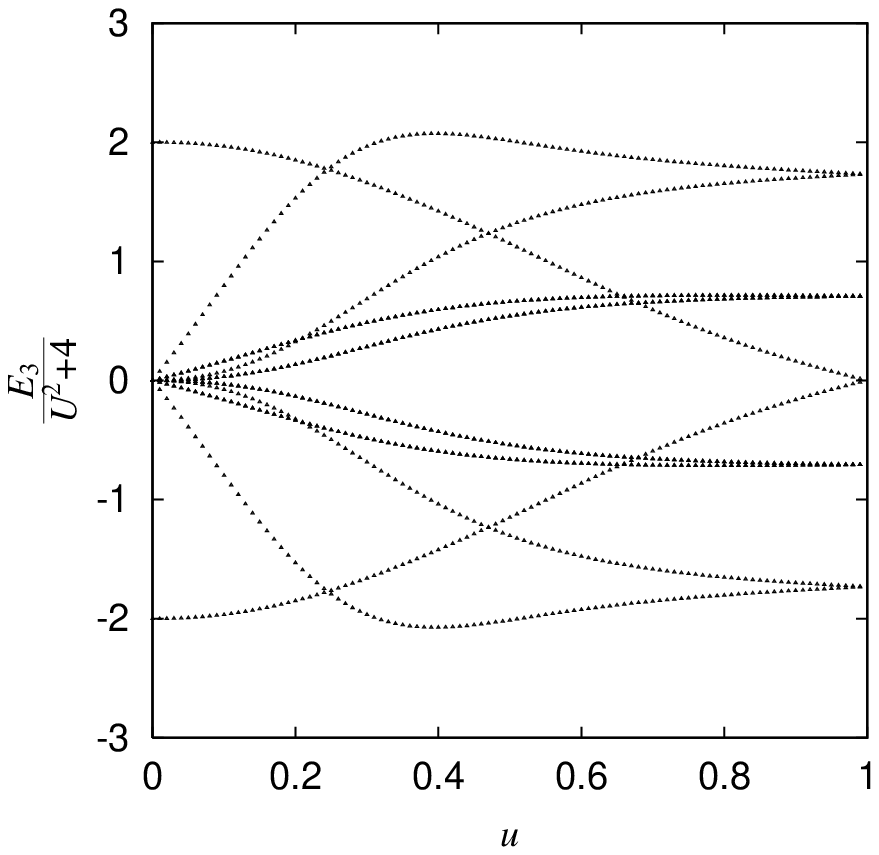}
\end{center}
\caption{Spectral flows $E_{3}$ from direct diagonalization
for $L=6, N=4, M=1$, $P=0$ and $S=\eta=1$.}
\label{fig:E3-641}
\end{figure}

We have to investigate the spectral flows of 
higher conserved operators.
Displayed in Figure \ref{fig:E3-641} and Table \ref{tab:T3-641}
are the spectral flows of the third conserved operator $I_{3}$
obtained by its direct diagonalization and its eigenvalues
at $u=0.3$ and $0.8$ given by the Bethe ansatz method, respectively.
Here we note that the vertical line in the figure indicates
the eigenvalue $E_{3}$ divided by $U^{2}+4$.
The eigenstates in a twofold permanent degeneracy 
in the energy spectral flows also permanently degenerate
in the spectral flows of $I_{3}$.
Figure \ref{fig:E3-641} and Table \ref{tab:T3-641} tell us that 
the eigenvalues $E_{3}$ of the eigenstates indexed by
$\{-\frac{5}{2},-\frac{3}{2},\frac{3}{2},\frac{5}{2};0;\}$ 
and $\{-\frac{1}{2},\frac{1}{2};;0\}$ never have degeneracies in $0<u<1$.
Hence they belong to the different eigenspaces of $I_{3}$
and the energy level crossing at $u=0.5$ is ``accounted for'' 
by the third dynamical symmetry.
Thus all the common eigenspaces of $\{H,I_{2},I_{3}\}$ 
with the set of quantum numbers 
$\{P=0,S=S_{z}=\eta=\eta_{z}=1\}$ are one-dimensional.

\begin{table}[p]
\begin{center}
\def\arraystretch{1.5}
\begin{tabular}{|l|l|l|l|}
\hline
 & \multicolumn{3}{|c|}{$E/(U+4)$} \\
\hline
 $\{I^{(i)};J;J^{\prime}\}$
 & \multicolumn{1}{|c|}{$u=0.4$} 
 & \multicolumn{1}{|c|}{$u=0.8$}
 & \multicolumn{1}{|c|}{$u=1.0$} \\
\hline
 $\{-\frac{5}{2},\frac{5}{2};;0\}$ 
 & $\;\;0.9497682678087262$ 
 & $\;\;0.5932037077099968$ 
 & $\;\;0.5$ \\
 $\{-\frac{5}{2},\frac{3}{2};;-1\}$ 
 & $\;\;0.5228330155542626$ 
 & $\;\;0.4784624958920988$ 
 & $\;\;0.5$ \\
 $\{-\frac{3}{2},\frac{5}{2};;1\}$ 
 & $\;\;0.5228330155542626$ 
 & $\;\;0.4784624958920990$ 
 & $\;\;0.5$ \\
 $\{-\frac{3}{2},\frac{3}{2};;0\}$ 
 & $\;\;0.3420419258315002$ 
 & $\;\;0.4083645387410650$ 
 & $\;\;0.5$ \\
 $\{-\frac{3}{2},\frac{1}{2};;-1\}$ 
 & $\;\;0.09563282828838615$ 
 & $\;\;0.3370068842393331$ 
 & $\;\;0.5$ \\
 $\{-\frac{1}{2},\frac{3}{2};;1\}$ 
 & $\;\;0.09563282828838634$ 
 & $\;\;0.3370068842393334$ 
 & $\;\;0.5$ \\
 $\{-\frac{5}{2},-\frac{3}{2},\frac{3}{2},\frac{5}{2};0;\}$ 
 & $\;\;0.0307824371531894$ 
 & $-0.2476848090979949$ 
 & $-0.5$ \\
 $\{-\frac{1}{2},\frac{1}{2};;0\}$ 
 & $-0.03078243715318940$ 
 & $\;\;0.2476848090979948$ 
 & $\;\;0.5$ \\
 $\{-\frac{5}{2},-\frac{3}{2},\frac{1}{2},\frac{5}{2};1;\}$ 
 & $-0.09563282828838634$ 
 & $-0.3370068842393332$ 
 & $-0.5$ \\
 $\{-\frac{5}{2},-\frac{1}{2},\frac{3}{2},\frac{5}{2};-1;\}$ 
 & $-0.09563282828838632$ 
 & $-0.3370068842393332$ 
 & $-0.5$ \\
 $\{-\frac{5}{2},-\frac{1}{2},\frac{1}{2},\frac{5}{2};0;\}$ 
 & $-0.3420419258315001$ 
 & $-0.4083645387410647$ 
 & $-0.5$ \\
 $\{-\frac{5}{2},-\frac{1}{2},\frac{1}{2},\frac{3}{2};1;\}$ 
 & $-0.5228330155542629$ 
 & $-0.4784624958920988$ 
 & $-0.5$ \\
 $\{-\frac{3}{2},-\frac{1}{2},\frac{1}{2},\frac{5}{2};-1;\}$ 
 & $-0.5228330155542631$ 
 & $-0.4784624958920992$ 
 & $-0.5$ \\
 $\{-\frac{3}{2},-\frac{1}{2},\frac{1}{2},\frac{3}{2};0;\}$ 
 & $-0.9497682678087262$ 
 & $-0.5932037077099968$ 
 & $-0.5$ \\
\hline
\end{tabular}
\def\arraystretch{1}
\end{center}
\caption{$E$ obtained by the Bethe ansatz method.}
\label{tab:T1-641}
\end{table}

\begin{table}[p]
\begin{center}
\def\arraystretch{1.5}
\begin{tabular}{|l|l|l|}
\hline
 & \multicolumn{2}{|c|}{$E_{2}/(U+4)$} \\
\hline
 $\{I^{(i)};J;J^{\prime}\}$
 & \multicolumn{1}{|c|}{$u=0.4$} 
 & \multicolumn{1}{|c|}{$u=0.8$} \\
\hline
 $\{-\frac{3}{2},-\frac{1}{2},\frac{1}{2},\frac{5}{2};-1;\!\}$\! 
 & $\;\;1.009981725805320$
 & $\;\;1.219335507531004$ \\
 $\{-\frac{3}{2},\frac{5}{2};;1\}$ 
 & $\;\;1.009981725805319$
 & $\;\;1.219335507531009$ \\
 $\{-\frac{5}{2},-\frac{3}{2},\frac{1}{2},\frac{5}{2};1;\!\}$\! 
 & $\;\;0.4769034635429986$
 & $\;\;1.068045373602690$ \\
 $\{-\frac{3}{2},\frac{1}{2};;-1\}$ 
 & $\;\;0.4769034635429980$
 & $\;\;1.068045373602692$ \\
 $\{-\frac{5}{2},-\frac{3}{2},\frac{3}{2},\frac{5}{2};0;\!\}$\! 
 & $\;\;0$
 & $\;\;0$ \\
 $\{-\frac{5}{2},-\frac{1}{2},\frac{1}{2},\frac{5}{2};0;\!\}$\! 
 & $\;\;0$
 & $\;\;0$ \\
 $\{-\frac{3}{2},-\frac{1}{2},\frac{1}{2},\frac{3}{2};0;\!\}$\! 
 & $\;\;0$
 & $\;\;0$ \\
 $\{-\frac{5}{2},\frac{5}{2};;0\}$ 
 & $\;\;0$
 & $\;\;0$ \\
 $\{-\frac{3}{2},\frac{3}{2};;0\}$ 
 & $\;\;0$
 & $\;\;0$ \\
 $\{-\frac{1}{2},\frac{1}{2};;0\}$ 
 & $\;\;0$
 & $\;\;0$ \\
 $\{-\frac{5}{2},-\frac{1}{2},\frac{3}{2},\frac{5}{2};-1;\!\}$\! 
 & $-0.4769034635429986$
 & $-1.068045373602690$ \\
 $\{-\frac{1}{2},\frac{3}{2};;1\}$ 
 & $-0.4769034635429986$
 & $-1.068045373602689$ \\
 $\{-\frac{5}{2},-\frac{1}{2},\frac{1}{2},\frac{3}{2};1;\!\}$\! 
 & $-1.009981725805320$
 & $-1.219335507531004$ \\
 $\{-\frac{5}{2},\frac{3}{2};;-1\}$ 
 & $-1.009981725805319$
 & $-1.219335507531006$ \\
\hline
\end{tabular}
\def\arraystretch{1}
\end{center}
\caption{$E_{2}$ obtained by the Bethe ansatz method.}
\label{tab:T2-641}
\end{table}

\begin{table}[p]
\begin{center}
\def\arraystretch{1.5}
\begin{tabular}{|l|l|l|}
\hline
 & \multicolumn{2}{|c|}{$E_{3}/(U^{2}+4)$} \\
\hline
 $\{I^{(i)};J;J^{\prime}\}$
 & \multicolumn{1}{|c|}{$u=0.4$} 
 & \multicolumn{1}{|c|}{$u=0.8$} \\
\hline
 $\{-\frac{5}{2},-\frac{3}{2},\frac{3}{2},\frac{5}{2};0;\}$
 & $\;\;2.073222007147403$
 & $\;\;1.802629509710799$ \\
 $\{-\frac{3}{2},\frac{3}{2};;0\}$
 & $\;\;1.421598188795819$
 & $\;\;0.3580668402927611$ \\
 $\{-\frac{5}{2},\frac{5}{2};;0\}$
 & $\;\;1.038079524262059$
 & $\;\;1.652488231566702$ \\
 $\{-\frac{5}{2},-\frac{3}{2},\frac{1}{2},\frac{5}{2};1;\}$
 & $\;\;0.5951023372314052$
 & $\;\;0.7147279328379515$ \\
 $\{-\frac{5}{2},-\frac{1}{2},\frac{3}{2},\frac{5}{2};-1;\}$
 & $\;\;0.5951023372314052$
 & $\;\;0.7147279328379515$ \\
 $\{-\frac{5}{2},\frac{3}{2};;-1\}$
 & $\;\;0.4301781122067006$
 & $\;\;0.6834852944297134$ \\
 $\{-\frac{3}{2},\frac{5}{2};;1\}$
 & $\;\;0.4301781122066977$
 & $\;\;0.6834852944296976$ \\
 $\{-\frac{5}{2},-\frac{1}{2},\frac{1}{2},\frac{3}{2};1;\}$
 & $-0.4301781122066989$
 & $-0.6834852944297066$ \\
 $\{-\frac{3}{2},-\frac{1}{2},\frac{1}{2},\frac{5}{2};-1;\}$
 & $-0.4301781122066986$
 & $-0.6834852944297066$ \\
 $\{-\frac{3}{2},\frac{1}{2};;-1\}$
 & $-0.5951023372314045$
 & $-0.7147279328379513$ \\
 $\{-\frac{1}{2},\frac{3}{2};;1\}$
 & $-0.5951023372314060$
 & $-0.7147279328379495$ \\
 $\{-\frac{3}{2},-\frac{1}{2},\frac{1}{2},\frac{3}{2};0;\}$
 & $-1.038079524262058$
 & $-1.652488231566708$ \\
 $\{-\frac{5}{2},-\frac{1}{2},\frac{1}{2},\frac{5}{2};0;\}$
 & $-1.421598188795818$
 & $-0.3580668402927629$ \\
 $\{-\frac{1}{2},\frac{1}{2};;0\}$
 & $-2.073222007147403$
 & $-1.802629509710803$ \\
\hline
\end{tabular}
\def\arraystretch{1}
\end{center}
\caption{$E_{3}$ obtained by the Bethe ansatz method.}
\label{tab:T3-641}
\end{table}

We give some comments on the results. 
 The origin of the twofold permanent degeneracies 
in Figure \ref{fig:E1-641} has not been clarified  
 until we obtain the correspondence 
between the energy spectral flows and the Bethe states.
Indeed we have found that they are due to the reflection symmetry.
At first sight, there is no crossing at the same value of $u$
in the spectral flows in Figure \ref{fig:E1-641}, 
\ref{fig:E2-641} and \ref{fig:E3-641}.
But we must not immediately conclude that
the energy level crossing in Figure \ref{fig:E1-641} 
is due to the dynamical symmetries 
since the spectral flows of $I_{2}$ and $I_{3}$ also include
permanent degeneracies.
Actually the eigenstates that have an energy level crossing 
in Figure \ref{fig:E1-641} 
permanently degenerate in the spectral flows of $I_{2}$
in Figure \ref{fig:E2-641}.
Thus the correspondence between the spectral flows and 
the Bethe states is crucial in the discussion above.

\section{Concluding remarks}

We have studied degeneracies in the energy spectrum of 
the one-dimensional Hubbard model with one down-spin on benzene.
The energy spectral flows have been obtained through 
the two approaches: direct diagonalization of the Hamiltonian matrix
and the Bethe ansatz method.
As is noted in~\cite{Heilmann-Lieb_71},
the former approach does not always determine the energy spectral flows 
since, if two of eigenvalues approach in numerical data, 
there are alternative spectral flows that may be drawn from the data.
To employ the latter, we have presented a procedure for giving 
numerical solutions of the Lieb--Wu equations for $UL>8$.
Combining the two approaches, we have shown that some energy spectral
flows have crossings which can not be understood by the
$U$-independent symmetries.
We have investigated the spectral flows of the higher conserved operators
$I_{2}$ and $I_{3}$ in the same way and
have found that the degenerate eigenstates are  
classified by the dynamical symmetries.

As we have indicated, 
the transfer matrix approach systematically
provides higher conserved operators.
Although it is not so easy to see their explicit form,
their eigenvalues can be written as functions of
the solutions for the Lieb--Wu equations~\eqref{eq:Lieb-Wu_eq}
by using the eigenvalues of the transfer 
matrix~\cite{Yue-Deguchi_97JPA,Ramos-Martins_97JPA}.
First five of them are given by
\begin{align}
 &\Tilde{E}_{n}=\sum_{i=1}^{N}e_{n}(k_{i}), \nn\\
 &e_{1}(k)=-2\cos k-\frac{1}{2}U, \nn\\
 &e_{2}(k)=-2\sin(2 k)-2U\sin k, \nn\\
 &e_{3}(k)=2\cos(3 k)+2\cos k+U\big(3\cos(2 k)-1\big)
  +\frac{3}{4}U^{2}\cos k-\frac{1}{16}U^{3}, \nn\\
 &e_{4}(k)=2\sin(4 k)+\frac{8}{3}\sin(2 k)
  +U \Big(4\sin(3 k)-\frac{4}{3}\sin k\Big)
  +2 U^{2}\sin(2 k), \nn\\
 &e_{5}(k)=-2\sin(5 k)-\frac{10}{3}\sin(3 k)-\frac{4}{3}\sin(3 k)
  -U\Big(5\cos k+\frac{1}{3}\Big) \nn\\
 &\quad\quad\quad
  -U^{2}\Big(\frac{15}{4}\cos(3 k)-\frac{5}{4}\cos k\Big)
  -U^{3}\Big(\frac{5}{8}\cos(2 k)+\frac{5}{24}\Big) 
  +\frac{5}{64}U^{4}\cos k-\frac{3}{256}U^{5}. \nn
\end{align}
Our numerical solutions for the Lieb--Wu 
equations~\eqref{eq:Lieb-Wu_eq} immediately gives their values
for $LU>8$.
We have also verified that these eigenvalues
never degenerate simultaneously.

Our studies on the spectral flows of conserved operators 
support several assumptions that 
have been believed for the Hubbard model and 
other Bethe-ansatz solvable models.
One is the one-to-one correspondence between solutions of
the Lieb--Wu equations and linearly independent eigenstates.
We have found that all the common eigenspaces of conserved operators 
$\{I_{n}\}$ are one-dimensional in several subspaces 
characterized by the sets of quantum numbers $\{N,M,P,S,\eta\}$,
which shows linear independence of the eigenvectors therein.
Another is the algebraic independence 
of the conserved operators $\{I_{n}\}$.
As is easily found, two commutative matrices which have 
the same type of degeneracy are not algebraically independent 
since one of the two matrices can be expressed by 
a linear combination of powers of another.
In our situations, the energy spectral flows have crossings 
at several values of $U$ which do not produce degeneracies
for the other conserved operators $I_{n}$.
This fact gives a necessary condition for the algebraic independence 
of $H$ and $I_{n}$.

We also remark that our discussion on the Lieb--Wu equations
in Section \ref{sec:LW-eq_one-down} give an evidence for 
the validity of the counting of states in Takahashi's string hypothesis.
It is interesting to generalize our procedure
to that with two or more down-spins,
which enables one to analyse the energy level crossings
in Heilmann and Lieb's situations~\cite{Heilmann-Lieb_71}.

\section*{Acknowledgements}
The authors would like to thank Prof.~M.~Wadati for helpful comments.
One of the authors (AN) appreciates the Research Fellowships of the
Japan Society for the Promotion of Science for Young Scientists. 
The present study is partially supported by the Grant-in-Aid for 
Encouragement of Young Scientists (A): No. 14702012.

\appendix

\section{Numerical solutions of Lieb--Wu equation}
\label{sec:numerical}

Following the procedure demonstrated in \ref{sec:LW-eq_one-down},
we give the numerical solutions $\{k_{i};\lambda\}$ of the Lieb--Wu 
equations~\eqref{eq:Lieb-Wu_eq} for $L=6, N=4, M=1, P=0$ and $S=\eta=1$.
Here, for the solutions indexed by
$\{I^{(i)};J;J^{\prime}\}=\{-I^{(i)};-J;-J^{\prime}\}$,
that is, $\{k_{i};\lambda\}=\{-k_{i};-\lambda\}$,
the Lieb--Wu equations~\eqref{eq:Lieb-Wu_eq} decouple as follows;
since $\sum_{i=1}^{4}k_{i}=0\, (\mathrm{mod}\, 2\pi)$ and $\lambda=0$,
the real wavenumber solutions can be given by 
$\{k_{1},k_{2},2\pi-k_{2},2\pi-k_{1};\lambda=0\}$ satisfying
\[
  \sin k_{i}=\frac{U}{4}\cot\big(3k_{i}\big),\quad (i=1,2),
\]
and the $k$-$\Lambda$-string solutions are
$\{k_{1},k_{2},\pi-\i\xi,\pi+\i\xi;\lambda=0\}$ satisfying
\[
  \sin k_{i}=\frac{U}{4}\cot\big(3k_{i}\big),\quad (i=1,2),\quad
  \sinh\xi=\frac{U}{4}\frac{\sinh\big(6\xi\big)}{\cosh\big(6\xi\big)-1}.
\]
We observe below that the solutions indexed by
$\big\{-\frac{5}{2},-\frac{3}{2},\frac{1}{2},\frac{5}{2};1;\big\}$
and $\{-\frac{1}{2},\frac{3}{2},\frac{5}{2};;1\}$ 
share the same $\lambda$. 
The similar situation is seen in other pairs of solutions.
Indeed they are in the relation of complementary solutions
by Woynarovich~\cite{Woynarovich_83JPC}, that is,
they give 8 (distinct) solutions of the 8th order algebraic equation
\[
  x^{6}(x^{2}-2\i(\lambda+\i U/4)-1)-(x^{2}-2\i(\lambda-\i U/4)-1)=0,
\]
through $x=\e^{\i k}$.

\vspace{5mm}
\noindent $\bullet\; u=0.4$
\def\arraystretch{1.1}
\begin{center}
\begin{tabular}{|l|lll|}
\hline
 $\{I^{(i)};J;J^{\prime}\}$
 & $k_{1}$
 & $k_{2}$ 
 & \\
 & $k_{3}$
 & $k_{4}$ 
 & $\lambda$ \\
\hline
 $\{-\frac{5}{2},-\frac{3}{2},\frac{3}{2},\frac{5}{2};0;\}$ 
 & $3.938730108910538$
 & $5.031918491061250$
 & \\
 & $1.251266816118337$
 & $2.344455198269048$
 & $10^{-16}$ \\
 $\{-\frac{5}{2},-\frac{3}{2},\frac{1}{2},\frac{5}{2};1;\}$ 
 & $4.051708143729197$
 & $5.108797294345751$
 & \\
 & $0.6064241499221595$
 & $2.799441026362065$
 & $0.7390790775734128$ \\
 $\{-\frac{5}{2},-\frac{1}{2},\frac{3}{2},\frac{5}{2};-1;\}$ 
 & $3.483744280817521$
 & $5.676761157257427$
 & \\
 & $1.174388012833835$
 & $2.231477163450389$
 & $-0.7390790775734128$ \\
 $\{-\frac{5}{2},-\frac{1}{2},\frac{1}{2},\frac{5}{2};0;\}$ 
 & $3.938730108910538$
 & $5.922043144971536$
 & \\
 & $0.3611421622080508$
 & $2.344455198269048$
 & $10^{-16}$ \\
 $\{-\frac{5}{2},-\frac{1}{2},\frac{1}{2},\frac{3}{2};1;\}$ 
 & $4.078300853771271$
 & $6.124165064590864$
 & \\
 & $0.7271767083756111$
 & $1.636727987621427$
 & $1.131437583776552$ \\
 $\{-\frac{3}{2},-\frac{1}{2},\frac{1}{2},\frac{5}{2};-1;\}$ 
 & $4.646457319558159$
 & $5.556008598803976$
 & \\
 & $0.1590202425887225$
 & $2.204884453408315$
 & $-1.131437583776552$ \\
 $\{-\frac{3}{2},-\frac{1}{2},\frac{1}{2},\frac{3}{2};0;\}$ 
 & $5.031918491061250$
 & $5.922043144971536$
 & \\
 & $0.3611421622080508$
 & $1.251266816118337$
 & $10^{-16}$ \\
\hline
\hline
 $\{I^{(i)};J;J^{\prime}\}$
 & $k_{1}$
 & $k_{2}$ 
 & \\
 & $\zeta$
 & $\xi$ 
 & $\lambda$ \\
\hline
 $\{-\frac{5}{2},\frac{5}{2};;0\}$ 
 & $3.938730108910538$
 & $2.344455198269048$ 
 & \\
 & $3.141592653589794$
 & $0.6481740825183825$ 
 & $-1\times 10^{-15}$ \\
 $\{-\frac{5}{2},\frac{3}{2};;-1\}$ 
 & $3.349805335791242$
 & $1.152236997712285$
 & \\
 & $4.032164140427823$
 & $0.9264247935487532$
 & $-1.131437583776552$ \\
 $\{-\frac{3}{2},\frac{5}{2};;1\}$ 
 & $5.130948309467302$
 & $2.933379971388344$
 & \\
 & $2.251021166751764$
 & $0.9264247935487532$
 & $1.131437583776552$ \\
 $\{-\frac{3}{2},\frac{3}{2};;0\}$ 
 & $5.03191849106125$
 & $1.251266816118337$
 & \\
 & $3.141592653589793$
 & $0.6481740825183825$
 & $10^{-16}$ \\
 $\{-\frac{3}{2},\frac{1}{2};;-1\}$ 
 & $4.833550063047541$
 & $0.2051607987234935$
 & \\
 & $3.763829876294069$
 & $0.7358326337961154$
 & $-0.7390790775734128$ \\
 $\{-\frac{1}{2},\frac{3}{2};;1\}$ 
 & $6.078024508456093$
 & $1.449635244132045$
 & \\
 & $2.519355430885517$
 & $0.7358326337961154$
 & $0.7390790775734128$ \\
 $\{-\frac{1}{2},\frac{1}{2};;0\}$ 
 & $5.922043144971536$
 & $0.3611421622080508$
 & \\
 & $3.141592653589793$
 & $0.6481740825183825$
 & $10^{-16}$ \\
\hline
\end{tabular}
\end{center}

\newpage

\noindent $\bullet\; u=0.8$
\def\arraystretch{1.2}
\begin{center}
\begin{tabular}{|l|lll|}
\hline
 $\{I^{(i)};J;J^{\prime}\}$
 & $k_{1}$
 & $k_{2}$ 
 & \\
 & $k_{3}$
 & $k_{4}$ 
 & $\lambda$ \\
\hline
 $\{-\frac{5}{2},-\frac{3}{2},\frac{3}{2},\frac{5}{2};0;\}$ 
 & $3.709764868570788$
 & $4.793788787327027$
 & \\
 & $1.489396519852559$
 & $2.573420438608797$
 & $10^{-16}$ \\
 $\{-\frac{5}{2},-\frac{3}{2},\frac{1}{2},\frac{5}{2};1;\}$ 
 & $3.950200892165781$
 & $5.005557244521997$
 & \\
 & $0.7483113471533719$
 & $2.862301130518022$
 & $3.876446955733767$ \\
 $\{-\frac{5}{2},-\frac{1}{2},\frac{3}{2},\frac{5}{2};-1;\}$ 
 & $3.420884176661565$
 & $5.534873960026214$
 & \\
 & $1.277628062657589$
 & $2.332984415013805$
 & $-3.876446955733767$ \\
 $\{-\frac{5}{2},-\frac{1}{2},\frac{1}{2},\frac{5}{2};0;\}$ 
 & $3.709764868570789$
 & $5.798257477867752$
 & \\
 & $0.4849278293118341$
 & $2.573420438608798$
 & $10^{-16}$ \\
 $\{-\frac{5}{2},-\frac{1}{2},\frac{1}{2},\frac{3}{2};1;\}$ 
 & $3.96270066619563$
 & $6.03997184302193$
 & \\
 & $0.7650646846850695$
 & $1.798633420456542$
 & $4.232098651896293$ \\
 $\{-\frac{3}{2},-\frac{1}{2},\frac{1}{2},\frac{5}{2};-1;\}$ 
 & $4.484551886723044$
 & $5.518120622494517$
 & \\
 & $0.2432134641576561$
 & $2.320484640983956$
 & $-4.232098651896293$ \\
 $\{-\frac{3}{2},-\frac{1}{2},\frac{1}{2},\frac{3}{2};0;\}$ 
 & $4.793788787327027$
 & $5.798257477867752$
 & \\
 & $0.4849278293118341$
 & $1.489396519852559$
 & $10^{-16}$ \\
\hline
\hline
 $\{I^{(i)};J;J^{\prime}\}$
 & $k_{1}$
 & $k_{2}$ 
 & \\
 & $\zeta$
 & $\xi$ 
 & $\lambda$ \\
\hline
 $\{-\frac{5}{2},\frac{5}{2};;0\}$ 
 & $3.70976486857079$
 & $2.573420438608798$ 
 & \\
 & $3.141592653589792$
 & $2.094719300551342$ 
 & $5\times 10^{-15}$ \\
 $\{-\frac{5}{2},\frac{3}{2};;-1\}$ 
 & $3.404556288703187$
 & $1.266195169893996$
 & \\
 & $3.947809577880995$
 & $2.454678139195706$
 & $-4.232098651896293$ \\
 $\{-\frac{3}{2},\frac{5}{2};;1\}$ 
 & $5.016990137285591$
 & $2.8786290184764$
 & \\
 & $2.335375729298592$
 & $2.454678139195706$
 & $4.232098651896293$ \\
 $\{-\frac{3}{2},\frac{3}{2};;0\}$ 
 & $4.793788787327028$
 & $1.48939651985256$
 & \\
 & $3.141592653589793$
 & $2.094719300551342$
 & $2\times 10^{-15}$ \\
 $\{-\frac{3}{2},\frac{1}{2};;-1\}$ 
 & $4.503384094618568$
 & $0.2564653496245172$
 & \\
 & $3.903260585058044$
 & $2.410923664087426$
 & $-3.876446955733767$ \\
 $\{-\frac{1}{2},\frac{3}{2};;1\}$ 
 & $6.026719957555069$
 & $1.779801212561019$
 & \\
 & $2.379924722121542$
 & $2.410923664087426$
 & $3.876446955733767$ \\
 $\{-\frac{1}{2},\frac{1}{2};;0\}$ 
 & $5.798257477867752$
 & $0.484927829311834$
 & \\
 & $3.141592653589793$
 & $2.094719300551341$
 & $10^{-16}$ \\
\hline
\end{tabular}
\end{center}


\end{document}